\documentclass[11pt]{article}

\usepackage{color}

\usepackage{amsmath, amssymb,amsfonts,longtable,tabularx}
\usepackage{cite}
\usepackage{colortbl}
\usepackage[all]{xy}
\usepackage{epsfig}
\usepackage{subfigure}
\usepackage{graphics}
\usepackage{multirow}

\usepackage{setspace}
\usepackage{caption}
\usepackage{pifont}
\usepackage[table]{xcolor}

\definecolor{gray1}{rgb}{0.9,0.9,0.9}
\definecolor{gray2}{rgb}{0.8,0.8,0.8}

\newcommand{\BigO}[1]{O\left(#1\right)}

\usepackage[hmargin=2cm, vmargin=2cm]{geometry}

\date{}

\usepackage{cite}

\begin{document}

\begin{flushleft}
{\huge
\textbf{The collapse of cooperation in evolving games}
}
\bigskip
\\
Alexander J. Stewart$^{1}$, 
Joshua B. Plotkin$^{1,2}$
\\
\bigskip
\bigskip
$^1$ Department of Biology, University of Pennsylvania, Philadelphia, PA 19104, USA
\\
$^2$ E-mail: jplotkin@sas.upenn.edu\\
\end{flushleft}

\vspace{1cm}


\subsection*{Abstract}

\noindent Game theory provides a quantitative framework for analyzing the
behavior of rational agents. The Iterated Prisoner's Dilemma in particular has
become a standard model for studying cooperation and cheating, with cooperation
often emerging as a robust outcome in evolving populations
\cite{Axelrod:1981kx,Axelrod2,Hofbauer:1998ys,Sig2,Nowak:2006ly,Nowak:2004fk,Rapoport:1965uq,Sigmund:2010ve}.  
Here we extend evolutionary game theory by allowing players' strategies as well as
their payoffs to evolve in response to selection on heritable mutations.  In
nature, many organisms engage in mutually beneficial interactions
\cite{Axelrod:2006zr,Axelrod2,Boyd:2010zr,Cordero:2012vn,Nowak:2006ys,Van-Dyken:2012ys,Waite:2012ly},
and individuals may seek to change the ratio of risk to reward for cooperation by
altering the resources they commit to cooperative interactions. To study this, we
construct a general framework for the co-evolution of strategies and payoffs in
arbitrary iterated games. We show that, as payoffs evolve, a trade-off between the
benefits and costs of cooperation precipitates a dramatic loss of cooperation
under the Iterated Prisoner's Dilemma; and eventually to evolution away from the
Prisoner's Dilemma altogether.  The collapse of cooperation is so extreme that the
average payoff in a population may decline, even as the potential payoff for
mutual cooperation increases.  Our work offers a new perspective on the Prisoner's
Dilemma and its predictions for cooperation in natural populations; and it
provides a general framework to understand the co-evolution of strategies and
payoffs in iterated interactions.

\clearpage
 
Iterated games provide a framework for studying social interactions
\cite{Axelrod:1981kx,Axelrod2,Hofbauer:1998ys,Nowak:2006ly,Nowak:2004fk,Rapoport:1965uq,Sigmund:2010ve},
allowing researchers to address pervasive biological problems such as the
evolution of cooperation and cheating
\cite{Axelrod:2006zr,Axelrod2,Boyd:2010zr,Cordero:2012vn,Nowak:2006ys,Van-Dyken:2012ys,Waite:2012ly}.
Simple examples such as the Prisoner's Dilemma and Snowdrift games
\cite{Akcay:2011ve,Hilbe:2013uq,Kummerli:2007qf,Maynard} showcase a startling
array of counter-intuitive social behaviors, especially when studied in a
population replicating under natural selection
\cite{Akin,Axelrod:1981kx,Imhof:2007uq,Maynard,Nowak:1993vn,Press:2012fk,Stewart:2013fk}.
Despite the subject's long history, a systematic treatment of all robust
evolutionary outcomes for even the simple Iterated Prisoner's Dilemma has only
recently emerged
\cite{Akin,Press:2012fk,Sig2,Stewart:2012ys,Stewart:2013fk}.

In an iterated two-player game, players $X$ and $Y$ face off in an infinite number
of successive ``rounds''.  In each round the players simultaneously choose their
plays and receive associated payoffs.   We study games with a $2\times2$ payoff
matrix, so that the players have two choices in each round.  We label these
choices ``cooperate'' ($c$) and ``defect'' ($d$), using the traditional language
for the Prisoner's Dilemma.  The four corresponding payoffs for player $X$ are
$\mathbf{R_x}=\left(R_x(cc), R_x(cd), R_x(dc),R_x(dd)\right)$, where $X$'s play is
denoted first.  $X$ may choose her play in each round depending on the
outcomes of all previous rounds \cite{Fund2,Traulsen:2006zr}. However, Press \& Dyson \cite{Press:2012fk} have
shown that a memory-1 player, whose choice depends only on the previous round, can
treat all opponents as though they also have memory-1. We therefore assume 
all players have memory-1 without loss of generality (see SI).  The strategy of
such a a player is described by the probabilities of cooperation given the four
possible outcomes of the previous round:
$\mathbf{p}=\left(p_{cc},p_{cd},p_{dc},p_{dd}\right)$.  The longterm average
payoff to player $X$ facing opponent $Y$, $s_{xy}$, can be calculated directly
from her strategy $\mathbf{p_{x}}$, her opponent's strategy $\mathbf{p_{y}}$, and
her payoffs $\mathbf{R_{x}}$. In the context of evolutionary game theory these
payoffs contribute to the player's fitness, which determines her expected
reproductive success in an evolving population (Figure~1).

\begin{figure}[h!] \centering \includegraphics[scale=1.0]{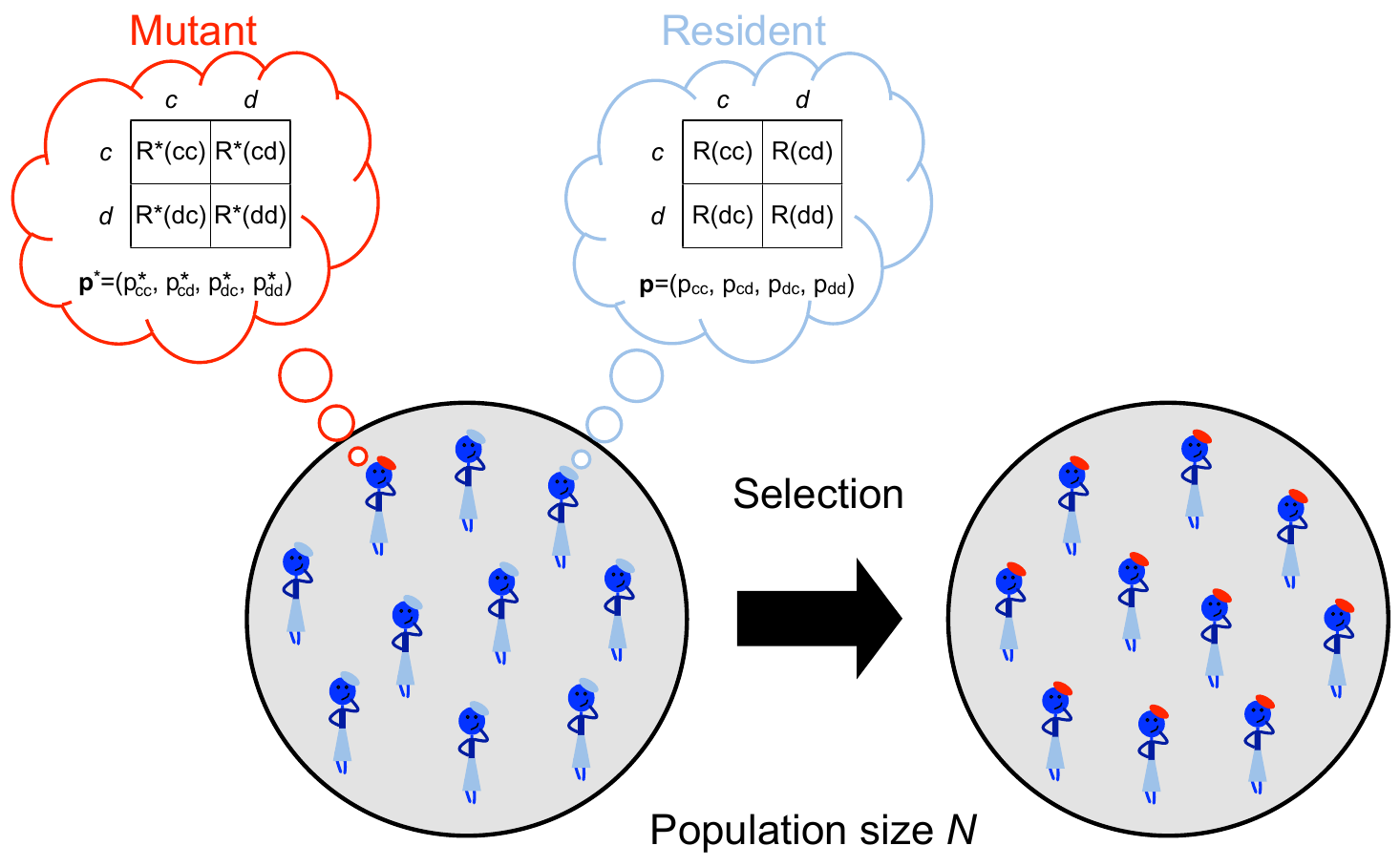}
\caption{\small Evolving the rules of the game. We model evolution in a population
of $N$ players who face each other in iterated, two-player games.  Each individual
has a ``genotype'' consisting of a strategy, $\mathbf{p}$, and a payoff matrix,
$\mathbf{R}$.  The payoffs received by a pair of players $X$ and $Y$ depends on
both players' strategies and payoff matrices.  Mutations are introduced
that change either a player's strategy or her payoff matrix.  Mutant strategies
are drawn uniformly from the four-dimensional space of memory-1 strategies.
Mutant payoff matrices are chosen according to a particular mutation scheme of
interest.  Natural selection and genetic drift occur according to a ``copying''
process \cite{Traulsen:2006zr}, in which two players, $X$ and $Y$, are selected at
random from the population, and $Y$ adopts the genotype of $X$ with probability
$f_{y \to x}=1/(1+\exp[\sigma(s_y-s_x)])$, where $s_x$ and $s_y$ are the 
payoffs the players receive in match-ups against the entire population, and
$\sigma$ is the strength of selection. Given a new mutant $X$ introduced into a
population with resident strategy $Y$, $X$ replaces $Y$ with probability
\cite{Traulsen:2006zr} $\rho=\left(\sum_{i=0}^{N-1}\prod_{j=1}^ie^{\sigma\left[(j-1)s_{yy}+(N-j)s_{yx}-js_{xy}-(N-j-1)s_{xx}\right]}\right)^{-1}$.  
} \end{figure}

The strategies that succeed in evolving populations can be understood in terms of
\textit{evolutionary robustness} \cite{Akin,Stewart:2013fk}.  A strategy is
evolutionary robust if, when resident in a population, no new mutant strategy is
favored to spread by natural selection.
Evolutionary robustness is a weaker condition than that of an evolutionary stable
strategy (ESS)\cite{Maynard,Maynard-Smith:1982vn} (see SI).  Robustness is a
useful notion because
there is rarely if ever a single ESS, as many strategies
$\mathbf{p_x}\neq\mathbf{p_y}$ are neutrally equivalent and can invade each other
by genetic drift. And so we focus on evolutionary robust strategies, which are
neutral to one another but resist invasion by any strategy outside of the set.
Indeed, the evolutionary robust strategies that cooperate amongst themselves are
already known to dominate in evolving populations playing the Iterated Prisoner's
Dilemma with standard payoffs \cite{Akin,Stewart:2013fk}.  

How does cooperation fare when both strategies and payoffs evolve in a population?
Here we expand the traditional purview of evolutionary game theory by allowing
heritable mutations to a players payoffs, as well as to their strategies, so that
the composition of payoffs and strategies in a population co-evolve over time. We
first consider evolution in the donation game, a form of Prisoner's Dilemma
\cite{Sig2,Stewart:2013fk} in which a player extracts a benefit $B$ if her
opponent cooperates and must pay a cost $C$ if she cooperates, resulting in
payoffs $R(cc)=B-C$, $R(cd)=-C$, $R(dc)=B$, and $R(dd)=0$. Each player has an
incentive to defect, although the players would receive a greater total payoff for
mutual cooperation.  It is natural to assume a trade-off so that mutations that
increase 
the benefit of cooperation, $B$, will also increase 
the cost of cooperation, $C$.  We therefore enforce the linear relation $B=\gamma
C + k$, and initially assume $\gamma>1$.

Starting from standard payoffs cooperation will quickly rise to high frequency in
populations when only strategies evolve (Figure~2a), as previously predicted
\cite{Hilbe:2013uq,Stewart:2013fk}.  But when both strategies and payoffs
co-evolve there is a striking reversal of fortunes. Evolution favors increasing
the benefits and costs of cooperation (Figure~2b), making the Prisoner's Dilemma
increasingly more acute over time. This evolution of the payoff matrix is
accompanied by a dramatic collapse of cooperation, so that the population is
eventually dominated by defection (Figure 2a).  Paradoxically, defection comes to
dominate even as the payoffs available for mutual cooperation continually increase
(Figure~2b).  Moreover, this collapse of cooperation is often accompanied by an
erosion of mean population fitness (Figure~2c).  

There is a simple intuition for this disheartening evolutionary outcome:
initially, the population is typically composed of self-cooperating strategies and
so mutations that increase the reward for mutual cooperation, $B-C$, are favored.
But such mutations also increase the ratio $B/(B-C)$, which is the temptation to
defect, to such a point that defection eventually out-competes cooperation.

\begin{figure}[h!] \centering \includegraphics[scale=1.0]{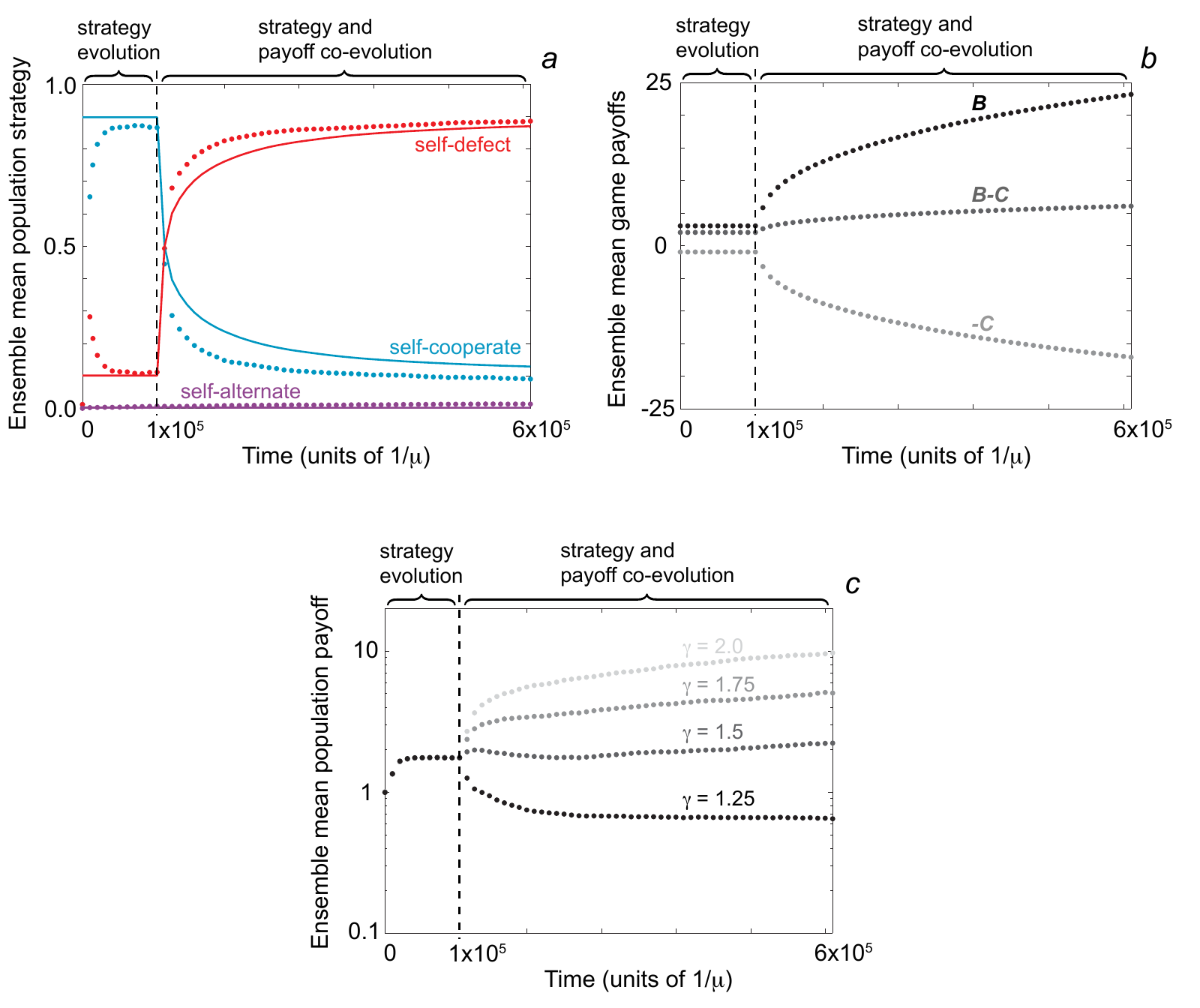}
\caption{\small The collapse of cooperation in the Prisoner's Dilemma.  We
simulated populations playing the iterated donation game, proposing mutant strategies
until reaching equilibrium, and then also proposing mutant payoffs, both at rate 
$\mu/2$.
Mutations to strategies were drawn 
uniformly from the full space of memory-1 strategies.  
Mutations to payoffs were drawn so that
increasing benefits of cooperation incur increasing costs: mutations perturbing
the benefit $B$ by $\Delta$ were drawn
uniformly from the range $\Delta\in\left[-0.1,0.1\right]$, with the corresponding
change to cost $C$ chosen to enforce the relationship $B=\gamma C + k$.
Evolution was modelled according to an imitation process under weak mutation \cite{Sig2,Stewart:2013fk,Traulsen:2006zr}.
(a) 
Cooperative strategies are
initially robust and dominate the population, but they are quickly replaced by defectors as payoffs evolve. 
Dots indicate the proportion of $10^5$ replicate simulated populations, at each time point, within distance
$\delta=0.01$ of the three strategy types self-cooperate, self-defect, and self-alternate. 
Lines indicate analytic predictions for the frequencies of these strategy
types, which depend upon the corresponding volumes of robust strategies (see SI).
(b) As payoffs evolve, the Prisoner's Dilemma becomes more acute, with both greater costs $C$ and benefits $B$
of cooperation.  Cooperation collapses even though the payoff for mutual cooperation, $B-C$, increases over time.  (c) The mean population
fitness (payoff) often declines over time, depending on the choice of parameter
$\gamma$. Populations of size $N=100$ were initiated with $B=3$ and $C=1$ (which determines $k$),
and they evolved under selection strength $\sigma=1$ (strong
selection), with $\gamma=1.25$ in panels a,b. } \end{figure}

We can understand the collapse of cooperation, and the co-evolution of strategies
and payoffs more generally, by determining which strategies are evolutionary
robust and how robustness varies as payoffs evolve. To do so, we have analytically
characterized all evolutionary robust memory-1 strategies for arbitrary $2\times2$
two-player iterated games (Figure~3, and SI).  In particular, we have proven the
following necessary condition: a robust memory-1 strategy must be one of three
types: self-cooperate, self-defect, or self-alternate. Self-cooperative strategies
$\mathcal{C}$ cooperate at equilibrium against an opponent using the same
strategy, meaning $p_{cc}=1$.  Conversely, self-defecting strategies $\mathcal{D}$
satisfy $p_{dd}=0$.  Self-alternating strategies $\mathcal{A}$ alternate between
cooperation and defection in subsequent rounds, meaning $p_{cd}=0$ and $p_{dc}=1$.
Monte-Carlo simulations on the full space of memory-1 strategies confirm
that populations adopt one of these three types $>97\%$ of the time, reflecting
the fact that all robust strategies fall within these three types. However, the robust
strategies are strict subsets of these types and, crucially, the volume of robust
strategies within each type depends on the payoffs of the game (Figure~3). The
robust volumes within these types can be computed analytically (see SI) and they
determine whether a population tends to adopt self-cooperation, self-defection, or
self-alternation (Figure~2a,b).

For the donation game illustrated in Figure~2, for example, the evolutionary robust strategies satisfy 
\begin{align*}
\nonumber &\mathcal{C}_{r}=\Big\{(p_{cc},p_{cd},p_{dc},p_{dd})| p_{cc}=1, p_{dc}<\frac{B}{C}(1-p_{cd}), p_{dd}<\left(\frac{B}{C}-1\right)(1-p_{cd})\Big\},\\
\nonumber &\mathcal{D}_{r}=\Big\{(p_{cc},p_{cd},p_{dc},p_{dd})| p_{dd}=0, p_{cc}<1-\left(\frac{B}{C}-1\right)p_{dc}, p_{cd}<1-\frac{B}{C}p_{dc}\Big\},\\
&\mathcal{A}_{r}=\Big\{(p_{cc},p_{cd},p_{dc},p_{dd})| p_{cd}=0, p_{dc}=1, p_{cc}<2\frac{C}{B+C}, p_{dd}<\frac{B-C}{B+C}\Big\},
\end{align*}
\\
for self-cooperating, self-defecting, and self-alternating strategies respectively. 
According to these equations, as the ratio $B/C$ decreases -- that is, as the
Prisoner's Dilemma becomes more acute --  the volume of the robust
self-cooperating strategies decreases whereas the volume of robust self-defecting
strategies grows. Thus it is the ratio of benefit to cost that matters for the
prospects of cooperation in the Iterated Prisoner's Dilemma, as payoffs and
strategies co-evolve.

\begin{figure}[h!] \centering \includegraphics[scale=0.7]{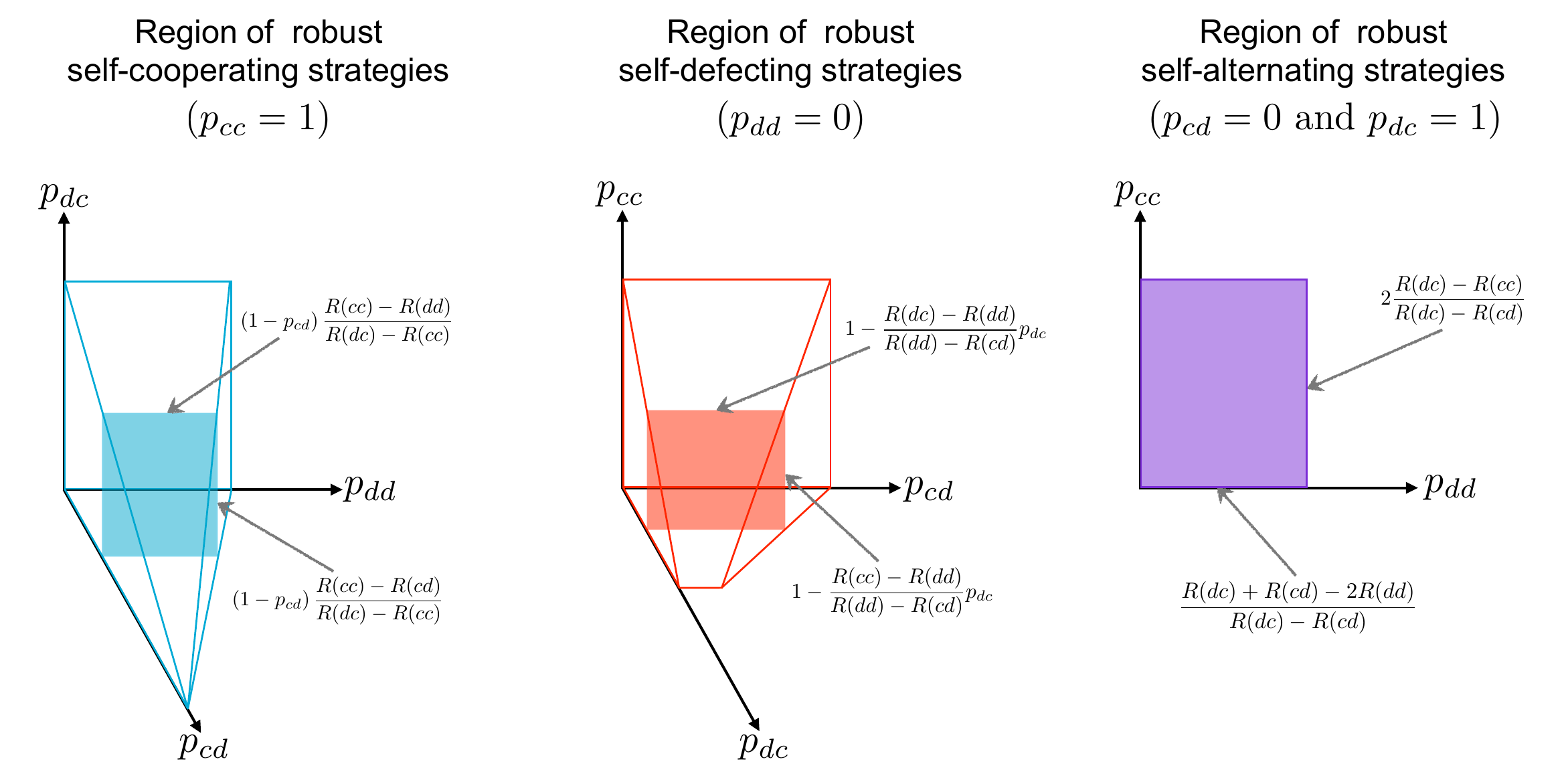}
\caption{\small Evolutionary robust strategies in iterated two-player games.
For an arbitrary $2\times2$ payoff matrix, an evolutionary robust memory-1 strategy must be one of three possible types:
those that cooperate against an opponent who cooperates (left, $p_{cc}=1$), those that defect against an opponent
who defects (center, $p_{dd}=0$), and those that alternate between cooperate and defect against an alternating opponent (right, $p_{cd}=0$ and $p_{dc}=1$). 
Within each of these strategy types, the strict subsets that are evolutionary robust can be determined analytically from the payoff matrix, as indicated 
on the figure (see SI for full derivations). The regions of robust self-cooperating and robust self-defecting strategies are three-dimensional, whereas the
robust self-alternating strategies are two-dimensional.
Monte Carlo simulations exploring the full space of memory-1 strategies confirm that these are the only evolutionary robust solutions (Fig.~S1).
As payoffs evolve in a population, the volumes of robust strategies change
according to the equations in the figure, and they determine the evolutionary dynamics
of cooperation and defection (Fig.~2).
} \end{figure}

Enforcing the payoff structure of the donation game restricts the population to a
Prisoner's Dilemma. However, our analysis applies to arbitrary $2\times2$ games
and mutation schemes, and so it can permit evolution between qualitatively
different types of games.  To explore this possibility, we considered a natural
generalization of the payoffs above by introducing another class of mutations that
enable a ``sucker" to recover some portion $\alpha$ of her lost benefit, resulting
in the payoff scheme: $R(cc)=B-C$, $R(cd)=-C+\alpha B$, $R(dc)=B$, and $R(dd)=0$.
As before, we allow the payoffs $B$ and $C$ to evolve together by enforcing a
linear relationship, and we additionally allow $\alpha \in [0,1]$ to evolve
independently.  This mutation scheme can produce any possible $2\times 2$ game.
In particular, when $\alpha<C/B$ the payoffs correspond to a Prisoner's Dilemma,
whereas when $\alpha>C/B$ the payoffs encode a Snowdrift game
\cite{Akcay:2011ve,Hilbe:2013uq,Kummerli:2007qf,Maynard}. 

Figure~4 shows the emergence of a qualitatively new game in a population
initialized at the Prisoner's Dilemma.  As payoffs and strategies co-evolve, the
benefits and costs of cooperation initially increase, resulting again in the
collapse of cooperation (Figure~4a). But this collapse is quickly followed by an
increase in $\alpha$ as the few remaining ``suckers" seek to recover payoffs lost
to defecting opponents. Eventually the Snowdrift game emerges (Figure~4b), and the
population is thereafter dominated by alternating strategies (Figure~4a).  The
instability of the Iterated Prisoner's Dilemma in favor of the Iterated Snowdrift
game is striking: $\alpha$ achieves its maximum value and the payoffs $B$ and
$C$ continually increase, producing increasingly acute versions of the Snowdrift
game.  

\begin{figure}[h!] \centering \includegraphics[scale=1.0]{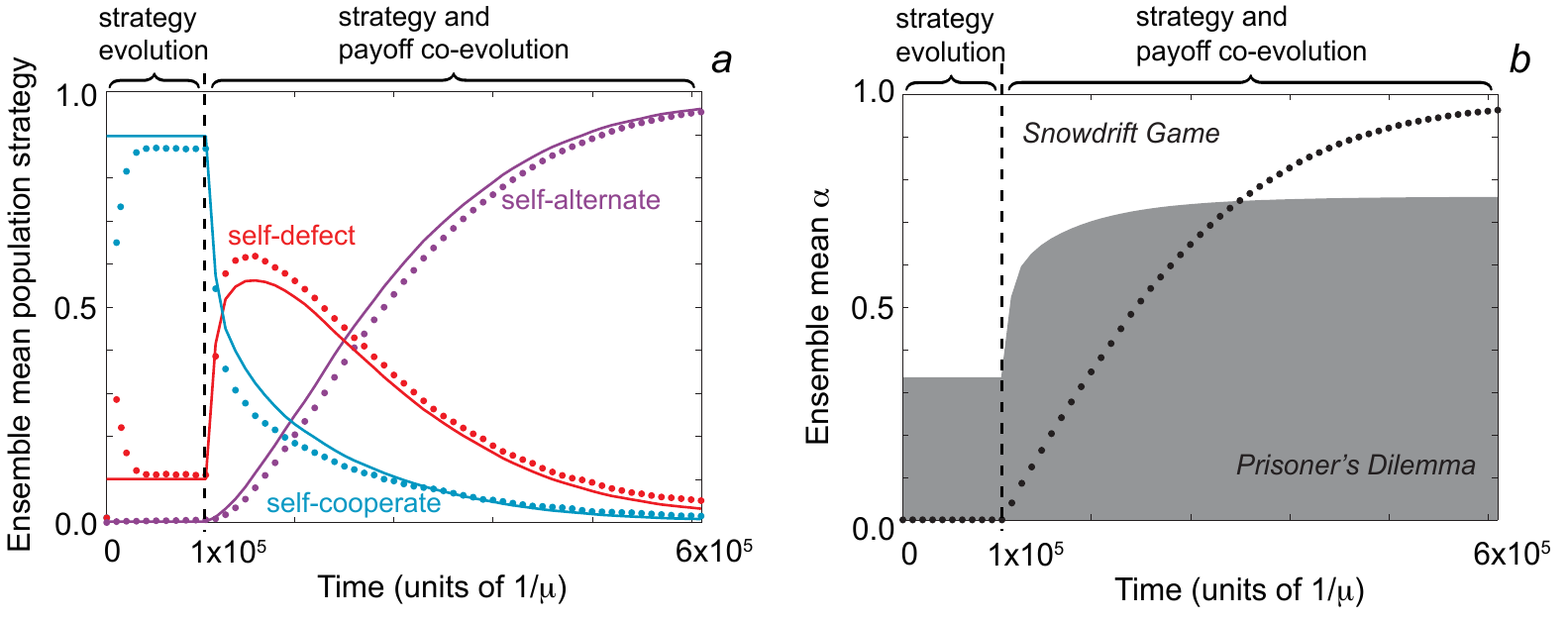}
\caption{\small Evolution from Prisoner's Dilemma to the Snowdrift game. 
We simulated a population under weak mutation, proposing mutant strategies drawn uniformly from the
full space of memory-1 IPD strategies. Aside from mutations to payoffs $B$ and $C$, as in Fig.~2, 
we also independently allowed mutations to $\alpha \in [0,1]$, so that ``suckers" can recover a portion $\alpha$ of the benefit lost
to a defecting opponent: $R(cd)=-C+\alpha B$.
(a) Evolution produces a rapid loss of cooperation and increase in defecting strategies, as in Fig.~2, now followed by an increase in alternating strategies.
Points indicate the proportion of simulated populations within a distance $\delta=0.01$ of
the three strategy types; lines indicate analytic predictions (see SI).
(b) Following the collapse of cooperation, the Prisoner's Dilemma (shaded region) is replaced by
the Snowdrift game (unshaded region), with $\alpha>C/B$. 
Parameters values as in Fig.~2a-b.
Populations of size $N=100$ were initiated with $B=3$ and $C=1$, and
evolved under selection strength $\sigma=1$ (strong
selection), with $\gamma=1.25$ 
} \end{figure}

We have assumed that payoff mutations are ``private", meaning that the player who
receives a payoff mutation unilaterally receives the increase or decrease in
benefit and cost. However our results are largely unchanged if we assume instead
that payoff mutations are ``public", such that a player sets her opponent's costs
(Figure~S5).  The collapse of cooperation also occurs under weak
selection (i.e.~$N \sigma \sim 1$; see SI), under local mutations to strategies
(Figure~S9), and also when mutations to payoffs are more rare than mutations to
strategies (Figure~S8). 

We have studied co-evolution of strategies and payoffs in a population of
individuals reproducing according to their finesses, which is the standard
framework used in evolutionary biology, as opposed to studying round-robin
tournaments \cite{Axelrod:1981kx,Axelrod2}, one-shot games \cite{Akcay:2011ve},
payoff evolution without strategy evolution \cite{Huang:2012fk,Huang:2012uq}, or
environmental ``shocks" to the payoff matrix \cite{Cabrales,Foster,Fund3}. We have
focused on payoff mutations that enforce a trade-off, by simultaneously increasing
the benefits and costs of cooperation.  However, our framework for analyzing
payoff-strategy co-evolution, based on computing the evolutionary robustness of
strategy sets, can be applied to any mutation scheme and can produce a potentially
vast array of evolutionary outcomes.  For example, if mutations increase the
benefit $B$ whilst leaving the cost $C$ unchanged then cooperation will remain
stable in the population (Figure~S3).  Alternatively, when mutations are private
and $0<\gamma<1$ (Figure~S4), so that costs increase faster than benefits, or when
mutations are public and players can set their opponent's benefits (Figure~S6),
then populations evolve towards an increasing frequency of self-cooperative
strategies despite decreased rewards for cooperation (Figure~S6). 

What types of payoff mutations can arise will depend upon the biological context.
Examples of the trade-off between costs and benefits that we have studied
(Figures~2 and
3) are found in nature at many scales \cite{Cordero:2012vn,Greig,Boyd:2010zr},
from human societies, where individuals modulate how frequently and how much they
punish free-riders \cite{Boyd:2010zr,Hauert:2007fk}, to micro-organism, such as
the marine bacteria Vibrionaceae \cite{Cordero:2012vn}. In Vibrionaceae
populations, for example, individuals cooperate in a public-goods game by sharing
siderophores required for iron acquisition. Mutations that alter the siderophore
biosynthetic pathway alter an individual's strategy, by changing its contribution
to the public good; mutations that improve the siderophore transport pathway alter
an individual's payoffs, by imposing a greater metabolic cost along with an
increased benefit from the public good. In these Vibrionaceae populations, as well
as many other biological systems with opportunities for cooperative interactions
\cite{Cordero:2012vn,Greig,Komarova:2012fk,Vulic:2001vn,Kerr:2002kx,Gore:2009uq},
defectors are often found at high frequency in nature, as our analysis predicts
when mutations increase both the costs and benefits of cooperation. A framework that
allows strategies and payoffs to co-evolve in a population significantly expands
the scope, and qualitatively alters the predictions, of evolutionary game theory
applied to biological systems.

\subsection*{Appendix}

In this supplement we prove that only three types of memory-1 strategies --
self-alternators, self-cooperators, and self-defectors -- can be evolutionary
robust and prevalent in arbitrary iterated two-player games with a $2\times2$
payoff matrix. We derive analytic expressions for the subsets of these
strategy types that are evolutionary robust. We show that the volume of robust
self-alternators, self-cooperators, and self-defectors provide good approximations
for the time spent by a population at each of these strategy types, for a fixed payoff matrix.  As shown in Fig.~2 and Fig.~4 of the main
text, this analysis enables us to explain strategy evolution in iterated
two-player games, even when payoff matrices are also allowed to evolve. In particular, this
analysis predicts the collapse of cooperation in the Iterated Prisoner's Dilemma (Fig.~2), as well as the transition from the Iterated Prisoner's
Dilemma to the Snowdrift game (Fig.~4).

We first define evolutionary robustness for an arbitrary iterated two-player game
with a $2\times2$ payoff matrix in a well-mixed population of finite size $N$; and
we state necessary and sufficient conditions for evolutionary robustness under the
limits of either strong or weak selection. We then show that, under strong
selection, only three subsets of memory-1 strategies can be evolutionary robust
and prevalent: self-cooperate, self-defect and self-alternate. Within each of
these three strategy types we derive the precise subset that is robust, and from
this we calculate the volume of robust strategies of each type.  Under weak
selection, by contrast, we show that only the self-cooperate and self-alternate
strategies can be evolutionary robust.

Finally we perform simulations under weak mutation for a variety of payoff mutation schemes, in addition to those used in the main text. These
simulations demonstrate that the volume of robust strategies within each of these types continues to determine the outcome of payoff-strategy
co-evolution in finite populations.

\subsection*{Iterated two-player games}

We consider an iterated game with an infinite number of successive rounds between two players, $X$ and $Y$. We study games with a $2\times2$ payoff
matrix, so that in each round each player has two choices, denoted cooperate $(c)$ or defect $(d)$. The payoffs for the respective players are given in Table
S1, in their most general form.


\begin{table*}[h]
\caption*{Table S1: Payoff matrix for an arbitrary $2\times2$ game}
\begin{center} 
\begin{tabular}{cccc}
&&\multicolumn{2}{c}{\textbf{Player} $\mathbf{Y}$} \\
& \multicolumn{1}{c|}{}&$c$ & $d$ \\ \cline{2-4}
&\multicolumn{1}{c|}{}&&\\
\multirow{3}{*}{\textbf{Player} $\mathbf{X}$}& \multicolumn{1}{c|}{$c$} &$R_x(cc),R_y(cc)$ &$R_x(cd),R_y(dc)$ \\
&\multicolumn{1}{c|}{}&&\\
&  \multicolumn{1}{c|}{$d$} &$R_x(dc),R_y(cd)$ &$R_x(dd),R_y(dd)$ \\
&\multicolumn{1}{c|}{}&&\\
\end{tabular}
 \end{center}
\end{table*}
 
\textbf{Memory:} In general, a player may have an arbitrarily long memory, such that her play in each round depends on the plays in all previous
rounds. However, as per Press and Dyson \cite{Press:2012fk}, a player with memory-1 may treat all opponents as though they are also memory-1,
regardless of the opponent's actual memory. And so the sets of scores $s_{xx}$, $s_{xy}$ and $s_{yx}$ for a player $X$ with memory-1 facing an opponent
$Y$ with arbitrary memory can be understood by considering the scores received by
$X$ against an arbitrary memory-1 opponent instead. By contrast, the
score a long-memory player $Y$ received against himself, $s_{yy}$, may depend on his memory capacity. Nonetheless, since our results for strong selection do not
depend on $s_{yy}$, we will show that a robust strategy for a memory-1 player $X$
is robust against any opponent, regardless of his memory capacity. The
same is true under weak selection: under
weak selection, the robustness of a strategy may depend on $s_{yy}$, but nevertheless, as shown in \cite{Stewart:2013fk} for the Prisoner's Dilemma, we can
still derive conditions for robustness under weak selection that do not depend on
the memory of $Y$. 

It is in this sense that our results on the evolutionary robustness of memory-1 strategies are without loss of generality -- because such memory-1 strategies are
evolutionary robust against all opponents, regardless of the opponent's memory
capacity. However, our results do not exlucde the possibility that there exist long-memory strategies that are also
evolutionary robust.
\\
\\ 
\textbf{Equilibrium payoffs in Iterated Games:} The longterm scores received by
two memory-1 players in an iterated two-player game are calculated
from the equilibrium rates of the different plays, $(cc)$, $(cd)$, $(dc)$ and $(dd)$, given by the stationary vector
$\mathbf{v}=(v_{cc},v_{cd},v_{dc},v_{dd})$ of the Markov matrix describing the iterated game \cite{Akin}. The equilibrium score of player $X$ against player
$Y$ is calculated according to

\begin{equation}
s_{xy}=\frac{\mathbf{v}\cdot\mathbf{R_x}}{\mathbf{v}\cdot\mathbf{I}}=\frac{D\left(\mathbf{p_x},\mathbf{p_y},\mathbf{R_x}\right)}{D\left(\mathbf{p_x},\mathbf{q_y},\mathbf{I}\right)}
\end{equation}
\\
where $\mathbf{I}=(1,1,1,1)$, $\mathbf{p_x}$ and $\mathbf{q_y}$ are the strategies of players $X$ and $Y$, and $\mathbf{R_x}$ is the payoff matrix of
player $X$. The determinant $D(\mathbf{p_x},\mathbf{q_y},\mathbf{f})$ gives the dot product between the stationary vector $\mathbf{v}$ and an
arbitrary vector $\mathbf{f}=(f_{cc},f_{cd},f_{dc},f_{dd})$ \cite{Press:2012fk}, where

\begin{equation}
D(\mathbf{p_x},\mathbf{q_y},\mathbf{f})=\det\left[ \begin{array}{cccc}
-1+p_{cc}q_{cc} & -1+p_{cc} & -1+q_{cc}& f_{cc} \\
p_{cd}q_{dc} & -1+p_{cd} & q_{dc}& f_{cd} \\
p_{dc}q_{cd} & p_{dc} & -1+q_{cd}& f_{dc} \\
p_{dd}q_{dd} & p_{dd} & q_{dd}& f_{dd}\\
 \end{array} \right].
\end{equation}
 \\
In general Eq.~1 is sufficient to calculate the scores received by a pair of memory-1 players. However, there are certain pathological cases in which the Markov chain
describing the iterated game has multiple absorbing states. The scores in these cases can be calculated by assuming that
players execute their strategy with some small ``error rate'' $\epsilon$ \cite{Fund2}, so that the probability of cooperation is at most $1-\epsilon$
and at least $\epsilon$. Assuming this, and taking the limit $\epsilon\to0$ then gives the player's scores in the cases where multiple absorbing
states exist.
\\
\\
\textbf{Alternate coordinate system:}
As shown in \cite{Akin,Press:2012fk,Stewart:2013fk}, manipulations of Eq.~1 produce an alternate coordinate system for the four-dimensional space of
memory-1 strategies, useful for analysing the outcomes of iterated two-player games. In particular, we can convert from the basis $(p_{cc},p_{cd},p_{dc},p_{dd})$ to the
basis $(\phi,\chi,\kappa,\lambda)$ \cite{Akin,Press:2012fk,Stewart:2013fk}, where the two coordinate systems are related by

\begin{equation}
\mathbf{\tilde{p}_x}=\phi\left[\mathbf{R_y}- \chi\mathbf{R_x}-(1-\chi)\kappa\mathbf{I}+\lambda\mathbf{L}\right].
\end{equation}
\\
Here $\mathbf{\tilde{p}_x}=(-1+p_{cc},-1+p_{cd},p_{dc},p_{dd})$, $\mathbf{I}=(1,1,1,1)$, and $\mathbf{L}=(0,1,1,0)$. 
To convert directly between the two coordinate systems we have the equations

\begin{eqnarray}
\nonumber p_{cc}&=&1-\phi\left(R_y(cc)-\chi R_x(cc)-(1-\chi)\kappa\right)\\
\nonumber p_{cd}&=&1-\phi\left(R_y(dc)-\chi R_x(cd)-(1-\chi)\kappa+\lambda\right)\\
\nonumber p_{dc}&=&\phi\left(\chi R_x(dc)-R_y(cd)+(1-\chi)\kappa-\lambda\right)\\
\nonumber p_{dd}&=&\phi\left((1-\chi)\kappa-R_y(dd)+\chi R_x(dd)\right).\\
\end{eqnarray}
\\
For completeness we have stated the coordinate transform above for the general case in which $\mathbf{R_x}\neq \mathbf{R_y}$.  Henceforth we will be
concerned with monomorphic populations in which  $\mathbf{R_x}= \mathbf{R_y}= \mathbf{R}$. In this coordinate scheme the players' scores are related
by \cite{Akin,Stewart:2013fk}

\begin{equation}
 s_{yx}-\chi s_{xy}-(1-\chi)\kappa+\lambda (v_{cd}+v_{dc})=0.
 \end{equation}
 \\
This relationship, which depends on the equilibrium rate of playing $(cd)$ and $(dc)$, can be used to determine analytic conditions for 
the evolutionary robust memory-1 strategies of an arbitrary $2\times2$ game. 
\\
\\
\textbf{Useful inequalities:} In addition to the relationship Eq.~5 we make note of four inequalities which we will use to determine the
memory-1 strategies that are evolutionary robust. We begin by noting that the equilibrium payoff for $X$ playing against an opponent $Y$ is given by
\cite{Akin}

\begin{equation}
s_{xy}=R(cc)v_{cc}+R(cd)v_{cd}+R(dc)v_{dc}+R(dd)v_{dd}
\end{equation}
\\
\\
(i) From  Eq.~6, the difference between the two players' scores can be written as
 
 \[
 s_{xy}-s_{yx}=(v_{dc}-v_{cd})(R_{dc}-R_{cd})
 \]
 \\
 which gives
 
 \begin{equation}
 s_{xy}-s_{yx}\leq (v_{cd}+v_{dc})|R_{dc}-R_{cd}|
 \end{equation}
\\
where equality is achieved by an opponent $Y$ for whom $v_{cd}=0$ (e.g.~an opponent who always cooperates).
\\
\\
(ii) Similarly, we must have

 \begin{equation}
 s_{xy}-s_{yx}\geq -(v_{dc}+v_{cd})|R_{dc}-R_{cd}|
 \end{equation}
\\
where equality is achieved by an opponent $Y$ for whom $v_{dc}=0$ (e.g.~an opponent who never cooperates).
\\
\\
(iii) From Eq.~6, the sum of the two players' scores is

\[
s_{xy}+s_{yx}=2(v_{cc}+(v_{dc}+v_{cd}))(R(cc)-R(dd))-(v_{dc}+v_{cd})(2R(cc)-(R(cd)+R(dc)))+2R(dd)
\]
\\
and, since $v_{cc}+(v_{dc}+v_{cd})\leq1$, we have

\begin{equation}
s_{xy}+s_{yx}\leq2R(cc)-(v_{dc}+v_{cd})(2R(cc)-(R(cd)+R(dc)))
\end{equation}
\\
where equality is achieved when $v_{dd}=0$ (e.g.~by an opponent who never defects once they have been defected against).
\\
\\
(iv) Finally, we also have

\begin{equation}
s_{xy}+s_{yx}\geq2R(dd)-(v_{dc}+v_{cd})(2R(dd)-(R(cd)+R(dc)))
\end{equation}
\\
where equality is achieved when $v_{cc}=0$ (e.g.~by an opponent who always defects once they have been cooperated with).

\subsection*{Evolution in a population of players}

We study evolution in a well-mixed, finite population of $N$ haploid, memory-1
players. Evolution is described by the ``imitation'' process of
\cite{Traulsen:2006zr}. Under this model, which is similar to the
Moran process, pairs of individuals, $X$ and $Y$, are drawn randomly from a
population of size $N$ at each time step. Player $X$ adopts the strategy of player
$Y$ with a probability $\left(1+\exp\left[\sigma(s_{x}-s_{y})\right]\right)^{-1}$
that depends on their respective total payoffs, $s_x$ and $s_y$, summed
across pairwise matchups with all players in the population.  Here $\sigma$
denotes the strength of selection.

We study evolution in the limit of weak mutation. This means that, at any point in time, the
population is monomorphic for some payoff matrix
$\mathbf{R}=(R(cc),R(cd),R(dc),R(dd))$ and some strategy
$\mathbf{p}=(p_{cc},p_{cd},p_{dc},p_{dd})$. Given a population monomorphic for the
resident type $X$, a mutation producing type $Y$ will fix with
probability \cite{Traulsen:2006zr}
\[
\rho(X,Y)=
\left(\sum_{i=0}^{N-1}\prod_{j=1}^ie^{\sigma\left[(j-1)s_{yy}+(N-j)s_{yx}-js_{xy}-(N-j-1)s_{xx}\right]}\right)^{-1},
\]
or otherwise will be lost.

The ``strong-selection" limit of this process is defined by taking $N\to\infty$
while keeping $\sigma$ fixed, so that even small differences in payoffs convey a
significant advantage to the player with the greater payoff.  Alternatively, the
``weak-selection" limit arises by taking $N\to\infty$ while keeping the product
$N\sigma\sim1$ fixed, in which case even deleterious strategies may reach high
frequency through genetic drift. We consider both of these regimes of selection in
our analyses below.

\subsection*{Evolutionary robustness}

We will use the above relations to determine which strategies are evolutionary robust in a population of $N$ players. 

The concept of evolutionary robustness\cite{Stewart:2013fk} is similar to the  notion of evolutionary stability
\cite{Maynard,Maynard-Smith:1982vn}.  An evolutionary stable strategy $\mathbf{p_x}$ is one that satisfies either $s_{xx}>s_{yx}$, or else
$s_{xx}=s_{yx}$ and $s_{xy}>s_{yy}$, for all opponents $ \mathbf{p_y} \neq \mathbf{p_x}$ \cite{Maynard,Maynard-Smith:1982vn}. This means that a
strategy is evolutionary stable provided (i) it cannot be selectively invaded by any other strategy ($s_{xx}>s_{yx}$), 
or (ii) it can selectively invade ($s_{xy}>s_{yy}$) any strategy
that can neutrally invade it ($s_{xx}=s_{yx}$).  However, as shown in \cite{Akin,Stewart:2013fk}, evolutionary stable strategies rarely exist within the full space of 
memory-1 strategies, because many strategies can neutrally invade each other. 
Therefore, we analyze the outcomes of evolution in a population using the notion of evolutionary robustness \cite{Stewart:2013fk}. 

In general, a strategy is defined to be evolutionary robust if, when resident in a population, there is no mutant that is favored to spread by natural selection
when rare \cite{Stewart:2013fk}. 
In particular, under strong selection a strategy $X$ is evolutionary robust selection iff, when resident in a population
of size $N$, it cannot be selectively invaded by any mutant $Y$ -- that is, iff $s_{xx}\geq s_{xy}$ for all $Y$. 
The condition for evolutionary robustness under strong selection is thus identical to that of a Nash equilibrium \cite{Akin}. 
Under weak selection, by contrast, a resident strategy $X$ is evolutionary robust iff the fixation
probability of any new mutant $Y$ satisfies $\rho_{yx}\leq 1/N$ (see \cite{Stewart:2013fk}). 
As shown previously \cite{Akin,Stewart:2013fk}, evolutionary robustness, as opposed to evolutionary stability, is useful for characterizing the
strategies that dominate in evolving populations.  In the remainder of the supplement we first derive results for evolutionary robustness under strong selection,
which are used in the main text.  We then derive conditions for evolutionary robustness under weak selection.

\subsection*{Necessary conditions for memory-1 strategies to be robust under strong selection}
We start by proving that strategies in the interior of the four-dimensional memory-1 strategy space cannot be evolutionary robust -- that is, they can always be
selectively invaded by some other strategy.  In fact, we will show that nothing on the interior can be robust with the exception of the ``equalizers'',
discussed below.

Consider a resident strategy $X$ characterised by $(\phi_x,\chi_x,\kappa_x,\lambda_x)$ and a mutant
strategy $Y$ characterised by $(\phi_y,\chi_y,\kappa_y,\lambda_y)$. From Eq.~5, with $y=x$, we find that the payoff of the resident against itself is

\[
s_{xx}=\kappa_{x}-\frac{\lambda_{x}}{1-\chi_{x}}(v_{cd}+v_{dc}).
\]
\\
Similarly, from Eq.~5 we find that the payoff of $Y$ against $X$ is

\[
s_{yx}=\frac{(1-\chi_x)\kappa_x-\lambda_x(w_{cd}+w_{dc})+\chi_x((1-\chi_y)\kappa_y-\lambda_y(w_{cd}+w_{dc}))}{(1-\chi_x \chi_y)}
\]
\\
where $\mathbf{v}$ is the stationary vector for $X$ playing against itself and $\mathbf{w}$ is the stationary vector for $X$ playing against $Y$.

First suppose that $X$ satisfies $0\leq p_{cc}<1$ and $0<p_{dd}\leq1$.
Then suppose $Y$ is chosen such that  $\chi_{x}=\chi_{y}$ and $\phi_x=\phi_y$. We can then write

\[
s_{yx}=\frac{(1-\chi_x)(\kappa_x+\chi_{x}\kappa_y)-(\lambda_x+\chi_x\lambda_y)(w_{cd}+w_{dc})}{(1-\chi_x ^2)}.
\]
\\
We also choose $\lambda_y$ such that $(1-\chi_x)\kappa_x-\lambda_x=(1-\chi_y)\kappa_y-\lambda_y$ i.e.~so that $p_{cd}$ and $p_{dc}$ are unaltered by the mutation. 
This gives

\[
s_{yx}=\frac{(\kappa_x+\chi_{x}\kappa_y)}{1+\chi_x}-\frac{\lambda_x}{1-\chi_x}(w_{cd}+w_{dc})-\frac{\chi_x(\kappa_y-\kappa_x)}{(1+\chi_x)}(w_{cd}+w_{dc})
\]
\\
We can then write

\[
s_{yx}-s_{xx}=\frac{\chi_{x}(\kappa_y-\kappa_x)}{1+\chi_x}\left[1-(w_{cd}+w_{dc})\right]+\frac{\lambda_{x}}{1-\chi_{x}}\left[(v_{cd}+v_{dc})-(w_{cd}+w_{dc})\right]
\]
\\
and so $Y$ selectively invades $X$ iff

\[
\frac{\chi_{x}(\kappa_y-\kappa_x)}{1+\chi_x}\left[1-(w_{cd}+w_{dc})\right]>\frac{\lambda_{x}}{1-\chi_{x}}\left[(w_{cd}+w_{dc})-(v_{cd}+v_{dc})\right]
\]
\\
This inequality can always be satisfied unless $v_{cd}+v_{dc}=1$, in which case both sides vanish and the mutation is neutral. To see this, we use Eq.~2 to calculate

\[
v_{cd}+v_{dc}=\frac{2 (1 - p_{cc}) (1 + p_{cc} - p_{dd}) p_{dd}}{(1 - p_{cc}) ((1 - p_{cd} - p_{dc})(1+p_{cc}) + 2 p_{cd} p_{dc} ) +  2 (1 - p_{cc}^2 + p_{cd} p_{dc}) p_{dd} - (1 - 2 p_{cc} + p_{cd} + p_{dc}) p_{dd}^2}
\]
\\
and assume that the mutant is such that $\kappa_{y}=\kappa_x+\eta$ where $\eta$ is small. We can then write

\begin{eqnarray*}
(w_{cd}+w_{dc})-(v_{cd}+v_{dc})=\\
(v_{cd}+v_{dc})\frac{\eta(1-(p_{cc}+p_{dd}))(1+p_{cc}-p_{dd})}{2 (1 - p_{cc}) (1 + p_{cc} - p_{dd}) p_{dd}}-\frac{\eta(1  - (p_{cc}+p_{dd}) +(p_{cc}-p_{dd})( p_{cd} + p_{dc}-(p_{cc}+p_{dd})))}{2 (1 - p_{cc}) (1 + p_{cc} - p_{dd}) p_{dd}}+\BigO{\eta^2}
\end{eqnarray*}
\\
or, more conveniently

\[
(w_{cd}+w_{dc})-(v_{cd}+v_{dc})=A\eta
\]
\\
where $A$ depends on the resident strategy and is finite (but can be zero). The condition for invasion of $X$ by $Y$ then becomes

\[
\frac{\chi_{x}}{1+\chi_x}\left[1-(v_{cd}+v_{dc})\right]\eta>\frac{\lambda_{x}}{1-\chi_{x}}A\eta
\]
\\
which can always be satisfied (since we can always invert the sign of $\eta$ by decreasing $\kappa_x$ instead of increasing it). The only exception
occurs when both sides of the inequality vanish, i.e.~if $w_{cd}+w_{dc}=v_{cd}+v_{dc}=1$, and $A=0$, or if $\chi_x=\lambda_x=0$ (we will deal will the latter case
separately).  Solving for $v_{cd}+v_{dc}=1$ gives solutions $p_{cd}=1,\ p_{dc}=0$ or $p_{cd}=0,\ p_{dc}=1$. Replacing these into the equation for
$w_{cd}+w_{dc}$ gives $A=0$. Therefore any strategy $X$ with $0<p_{cc}<1$ and $0<p_{dd}<1$ can be selectively invaded unless $p_{cd}=1$ and
$p_{dc}=0$, or $p_{cd}=0$ and $p_{dc}=1$ -- that is, unless $X$ alternates.  

In the boundary cases $p_{cc}=0$ or $p_{dd}=1$, $\phi$ takes a maximal value, and an increase in $\kappa$
necessitates a corresponding decrease in $\phi$ to ensure all probabilities are in the range $[0,1]$. However, the same argument holds as above,
since small changes in $\kappa$ and $\phi$ together precipitate a small change in $v_{cd}+v_{dc}$; so that even in these boundary cases, $X$ can be
selectively invaded unless  $p_{cd}=1$ and $p_{dc}=0$,  or $p_{cd}=0$ and $p_{dc}=1$ -- that is, unless $X$ alternates.

The self-alternating strategies discussed so far come in two forms. However, we can show that the alternating strategies of the form
$p_{cd}=1$ and $p_{dc}=0$ cannot be robust. It is easy to construct a mutant that can selectively invade such strategies: if we assume, without loss of generality, that
$R(dc)>R(cd)$, then an opponent with $p_{dc}=0$ and $p_{cd}<1$ scores $R(dc)$ at equilibrium, whereas such an alternator scores $(1/2)(R(cd)+R(dc))$ against
itself. Therefore the mutant can selectively invade and this strategy type cannot be robust. The only exception occurs in the special case
$R(cd)=R(dc)$, in which case both types of alternators score identically. Henceforth we consider only alternators of the form $p_{cd}=0$ and $p_{dc}=1$, 
because only these alternators have the potential to be evolutionary robust.
\\
\\ 
Now we consider the cases $X$ that satisfy $p_{cc}=1$ or $p_{dd}=0$, that is the self-cooperators and self-defectors. 
To address these cases, we can use the same procedure as above. 
We consider a mutant $Y$ such that $\kappa_x=\kappa_y$, $\chi_x=\chi_y$ and $\phi_x=\phi_y$. This has the
effect that $p_{cc}$ and $p_{dd}$ remain constant under mutation. We then have

\[
s_{yx}=\kappa_x-\frac{(\lambda_x+\chi_x\lambda_y)(w_{cd}+w_{dc})}{(1-\chi_x ^2)}
\]
\\
and $Y$ can selectively invade iff

\[
\lambda_{x}\left[(v_{cd}+v_{dc})-(w_{cd}+w_{dc})\right]>\chi_x\left[\lambda_y(w_{cd}+w_{dc})-\lambda_x(v_{cd}+v_{dc})\right]
\]
\\
As in the previous case, it is easy to show that a small change $\eta$ in $\lambda_x$ gives $(w_{cd}+w_{dc})-(v_{cd}+v_{dc})=A\eta$ 
where $A$ depends on the resident strategy. We then have

\[
-(1+\chi_x)A\eta\lambda_{x}>\eta\chi_x(v_{cd}+v_{dc})
\]
\\
Once again, this can always be satisfied by either increasing or decreasing $\lambda_x$. The only exception occurs if both sides vanish so that the mutant
is neutral, i.e.~if
$w_{cd}+w_{dc}=v_{cd}+v_{dc}=0$ and $A=0$, or $\lambda_x=\chi_x=0$ (called equalizers, see below). The former case occurs iff either (i) $X$ is a
self-cooperator with $p_{cc}=1$, (ii) $X$ is a self-defector with $p_{dd}=0$, or (iii) $X$ satisfies $p_{cc}=0$ and $p_{dd}=1$.
However the case $p_{cc}=0$ and $p_{dd}=1$ can be selectively invaded, as shown above, and it is therefore not robust.

Therefore, in total, we have proven that any evolutionary robust strategy for an arbitrary $2\times2$ game must be one of following four types:
\begin{itemize}
\item the self-cooperators $\mathcal{C}=\{\left(p_{cc},p_{cd},p_{dc},p_{dd}\right)|p_{cc}=1\}$, 
\item the self-defectors $\mathcal{D}=\{\left(p_{cc},p_{cd},p_{dc},p_{dd}\right)|p_{dd}=0\}$, 
\item the self-alternators $\mathcal{A}=\{\left(p_{cc},p_{cd},p_{dc},p_{dd}\right)|p_{cd}=0,\ p_{dc}=1\}$, 
\item the equalizers $\mathcal{E}=\{(\phi,\chi,\kappa, \lambda)|\lambda=\chi=0\}$.
\end{itemize}


Finally, any strategy that falls in the intersection of two types above (e.g.~those satisfying both $p_{cc}=1$ and $p_{dd}=0$) cannot be robust. Such a
``mixed-type'' strategy $X$ will receive a score $s_{xx}$ that is a linear combination of the scores received by a ``pure-type'' strategy. However whichever 
pure-type strategy receives the higher of the two scores against itself can invade such a ``mixed-type'' strategy. Nonetheless, such ``mixed-type'' strategies 
spread neutrally in a population of robust strategies with which they share an absorbing state.

\subsection*{Necessary and sufficient conditions for evolutionary robustness under strong selection}
As discussed above, a memory-1 strategy that is evolutionary robust in an arbitrary $2\times2$ game must belong to one of the four types: self-alternators, self-cooperators, 
self-defectors,
or equilizers. We now derive sufficient conditions for strategies of each of these types to be robust.
\\
\\
\textbf{The self-cooperators:}
The self-cooperators $\mathcal{C}$ satisfy $p_{cc}=1$ and score $s_{xx}=R(cc)$ against themselves, which corresponds to $\kappa=R(cc)$. In the context of the Iterated Prisoner's Dilemma, these are precisely the ``Good'' strategies of  \cite{Akin} and discussed it \cite{Stewart:2013fk}. In order to invade a resident
strategy $X$, a mutant $Y$ must have

\[
s_{yx}>R(cc).
\]
\\
Combining this with Eq.~5, Eq.~9, and Eq.~10, and rearranging, we find that $Y$ can selectively invade iff

\[
-\chi(R(cc)-(R(cd)+R(dc))>\lambda
\]
\\
or

\[
-\chi(R(dc)-R(cd))>\lambda.
\]
\\
Converting back to our original coordinate system, this implies a self-cooperator $X$ is robust iff:

\[
p_{dc}(R(dc)-R(cc))<(R(cc)-R(cd))(1-p_{cd})
\]
\\
and
\begin{equation}
p_{dd}(R(dc)-R(cc))<(R(cc)-R(dd))(1-p_{cd}).
\end{equation}
\\
The evolutionary robust self-cooperating strategies are thus described by the set
\[
\mathcal{C}_{r}=\left\{\mathbf{p} \ | \ p_{cc}=1,p_{dc}<\frac{R(cc)-R(cd)}{R(dc)-R(cc)}(1-p_{cd}),p_{dd}<\frac{R(cc)-R(dd)}{R(dc)-R(cc)}(1-p_{cd})\right\}.
\]
\\
These analytic expressions for the robust self-cooperating strategies are confirmed by Monte-Carlo simulations (Fig.~S1).
\\
\\
\textbf{The self-defectors:}
The self-defectors $\mathcal{D}$ satisfy $p_{dd}=0$ and score $s_{xx}=R(dd)$ against themselves, which corresponds to $\kappa=R(dd)$. In order to invade, a mutant $Y$ must therefore have
\[
s_{yx}>R(dd).
\]
\\
using this, as well as Eq.~5, Eq.~7 and Eq.~9, we find that a strategy $Y$ can invade iff

\[
\chi(R(dc)-R(cd))<\lambda
\]
\\
or

\[
\chi(R(cd)+R(dd)-2R(dd))<\lambda
\]
\\
Converting back to our original coordinate system, this implies that a self-defector $X$ is robust iff:

\[
p_{dc}(R(cc)-R(dd))<(R(dd)-R(cd))(1-p_{cc})
\]
\\
and
\begin{equation}
p_{dc}(R(dc)-R(dd))<(R(dd)-R(cd))(1-p_{cd}).
\end{equation}
\\
The evolutionary robust self-defecting strategies are thus described by the set
\[
\mathcal{D}_{r}=\left\{\mathbf{p} \ | \ p_{dd}=0,p_{dc}<\frac{R(dd)-R(cd)}{R(cc)-R(dd)}(1-p_{cc}),p_{dc}<\frac{R(dd)-R(cd)}{R(dc)-R(dd)}(1-p_{cd})\right\}.
\]
\\
These analytic expressions for the robust self-defecting strategies are confirmed by Monte-Carlo simulations (Fig.~S1).
\\
\\
\textbf{The self-alternators:}
The self-alternators $\mathcal{A}$ satisfy  $p_{cd}=0$ and $p_{dc}=1$. 
Using Eq.~4, and converting to the alternate coordinate system, we have

\[
\lambda=(1-\chi)\left(\kappa-\frac{R(cd)+R(dc)}{2}\right)
\]
\\
for strategies of this type.
From Eq.~5, a resident strategy $X$ of this type has
\[
s_{xx}=\frac{R(cd)+R(dc)}{2}.
\]
\\
In order to selectively invade the resident, then, a mutant $Y$ must  satisfy
\[
s_{yx}>\frac{R(cd)+R(dc)}{2}
\]
\\
Combining this with Eq.~5, Eq.~9, and Eq.~10, and rearranging, we find that $Y$ can selectively invade iff

\[
(1+\chi)\left(\frac{R(cd)+R(dc)}{2}-\kappa\right)<2\chi(R(cc)-\kappa)
\]
\\
or

\[
(1+\chi)\left(\frac{R(cd)+R(dc)}{2}-\kappa\right)<2\chi(R(dd)-\kappa)
\]
\\
Converting back to our original coordinate system, this implies an self-alternator $X$ is robust iff:
\[
p_{cc}<2\frac{R(dc)-R(cc)}{R(dc)-R(cd)}
\]
\\
and
\begin{equation}
p_{dd}<\frac{R(dc)+R(cd)-2R(dd)}{R(dc)-R(cd)}.
\end{equation}
\\
The evolutionary robust self-alternating strategies are thus described by the set
\[
\mathcal{A}_{r}=\left\{\mathbf{p} \ | \ p_{cd}=0,p_{dc}=1,p_{cc}<2\frac{R(dc)-R(cc)}{R(dc)-R(cd)},p_{dd}<\frac{R(dc)+R(cd)-2R(dd)}{R(dc)-R(cd)}\right\}.
\]
\\
These analytic expressions for the robust self-alternating strategies are confirmed by Monte-Carlo simulations (Fig.~S1).
\\
\\
\textbf{Characteristics of evolutionary robust strategies:} We have identified the robust subsets of self-alternators, self-cooperators and self-defectors,
$\mathcal{A}_{r}$, $\mathcal{C}_{r}$ and $\mathcal{D}_{r}$, which cannot be selectively invaded, under strong selection. The
inequalities Eqs.~11-13 defining these robust strategies in fact guarantee that any invading strategy is selected against, unless it satisfies $w_{cd}+w_{dc}=1$
in the case of self-alternators, or $w_{cd}+w_{dc}=0$ in the case of self-cooperators and self-defectors -- in which case both sides of the inequality vanish and the
mutant invades neutrally. The mutant strategies that satisfy these conditions are precisely those of the same type (self-alternator, self-cooperator or
self-defector) as the resident.  In other words, the only strategies that can neutrally invade robust self-alternators are other self-alternating strategies; and
the only strategies that can neutrally invade robust self-cooperators are other self-cooperators; and the only strategies that can neutrally invade robust
self-defectors are other self-defectors. 
\\
\\
\textbf{The Equalizers:} Finally, we must deal with the case of the Equalizers \cite{Boer}, which have $\chi=\lambda=0$.
From Eq.~5, we see that such strategies satisfy $s_{yx}=\kappa$ against any invader $Y$.  Thus, 
a population of Equalizers is neutral against all possible invaders. The equalizer strategies are thus evolutionary robust. However, 
unlike the other sets of robust strategies ($\mathcal{C}_r,\ \mathcal{D}_r,\ \mathcal{A}_r$), which resist replacement by any other strategy type, equalizers
never resist invasion, and so they tend to be quickly lost from a population through neutral drift. Therefore we exclude them from our further discussion of
robust strategies and, indeed, we find that populations spend very little time ($<0.01\%$) at the equalizers.
\\
\\
\textbf{Volume of a robust strategy type:}
We can use Eqs.~11-13 to calculate the volumes associated with each robust strategy type. 
In the case of the self-alternators the volume of $\mathcal{A}_{r}$ is in fact a 2D surface of area

\[
\left( 2\frac{R(dc)-R(cc)}{R(dc)-R(cd)}\right)\times\left(\frac{R(dc)+R(cd)-2R(dd)}{R(dc)-R(cd)}\right)
 \]
 \\
where, in addition, we must constrain the area so that only strategies within the unit square are included.
Similarly, $\mathcal{C}_{r}$ has cross-sections of area

\[
\left(\frac{R(cc)-R(cd)}{R(dc)-R(cc)}(1-p_{cd})\right)\times\left(\frac{R(cc)-R(dd)}{R(dc)-R(cc)}(1-p_{cd})\right)
\]
\\
and its volume is calculated by integration, with the limits of integration chosen to include only strategies lying within the unit cube.
Finally, $\mathcal{D}_{r}$ has cross-sections of area

\[
\left(1-\frac{R(cc)-R(dd)}{R(dd)-R(cd)}p_{dc}\right)\times\left(1-\frac{R(dc)-R(dd)}{R(dd)-R(cd)}p_{dc}\right)
\]
\\
and its volume is calculated by integrating across those strategies lying within the unit cube.
\\
\\
\textbf{Time spent at different strategy types:}
We now use our results on the volumes of robust strategies to approximate the time spent at the different strategy types -- self-cooperators, self-defectors, and
self-alternators -- for fixed payoffs under strong selection.
To make this analytical approximation we will assume that the population spends all of its time at these three strategy types, 
an approximation motivated by the fact that these types contain all the evolutionary robust strategies (except for the equilizers, which are quickly replaced
through neutral drift). Indeed, Monte Carlo simulations confirm that populations spend $>97\%$ of their time at self-alternators, self-cooperators or self-defectors, 
for values of payoffs ranging across an order of magnitude. 

To approximate the amount of time a population spends in $\mathcal{C}$,
$\mathcal{D}$ or $\mathcal{A}$, we simply the evolution of strategies in population as a
three-state Markov chain (Fig.~S2). We assume that the probability $g$ of entering
a strategy type is given by the probability that a robust strategy of that type
replaces a randomly drawn memory-1 strategy. We assume that non-robust strategies can be
neglected, because although they may be able to invade, they can quickly be
reinvaded. 

In order to calculate the probability of enetering a strategy type under the ``imitation'' model of \cite{Traulsen:2006zr}, we use the probability that a new
mutant $Y$ fixes in a population otherwise comprised of a resident $X$:

\[
\rho(\mathbf{p_x},\mathbf{q_y})=
\left(\sum_{i=0}^{N-1}\prod_{j=1}^ie^{\sigma\left[(j-1)s_{yy}+(N-j)s_{yx}-js_{xy}-(N-j-1)s_{xx}\right]}\right)^{-1}
\]
\\
The probability of the population adopting a self-alternator strategy under
in this three-state chain is
then

\[
g_{a}=Z\delta^2V_a\int_{\mathbf{p}\in[0,1]^4}\int_{\mathbf{q}\in\mathcal{A}_{r}}\rho(\mathbf{p},\mathbf{q})\mathbf{dp}\mathbf{dq}
\]
\\
where $\mathbf{q}$ is integrated over the set of robust self-alternating strategies, $\mathbf{p}$ is integrated over the full set of memory-1 strategy,
$\rho(\mathbf{p},\mathbf{q})$ is the probability that a resident strategy $\mathbf{p}$ is replaced by a robust alternator $\mathbf{q}$,
and $V_A$ is the two-dimensional area comprised by robust alternators. The term 
$\delta^2V_a$ denotes the volumes of all memory-1 strategies within Euclidean distance $\delta$ of the robust alternators, called the 
$\delta$-neighborhood of the robust alternators \cite{Sig2,Stewart:2013fk}. 
The constant term $Z$ normalizes the probability of adopting a strategy, so that $g_{a}+g_{c}+g_{d}=1$. 

Similarly, the probability of the
system adopting a robust self-cooperator is

\[
g_{c}=Z\delta V_c\int_{\mathbf{p}\in[0,1]^4}\int_{\mathbf{q}\in\mathcal{C}_{r}}\rho(\mathbf{p},\mathbf{q})\mathbf{dp}\mathbf{dq},
\]
\\
and the probability of the system adopting a robust self-defector strategy

\begin{equation}
g_{d}=Z\delta V_d\int_{\mathbf{p}\in[0,1]^4}\int_{\mathbf{q}\in\mathcal{D}_{r}}\rho(\mathbf{p},\mathbf{q})\mathbf{dp}\mathbf{dq}.
\end{equation}
\\
Once at a robust strategy, we know that, under strong selection, the system evolves through neutral invasion among strategies of the same type ($\mathcal{C}$, $\mathcal{D}$, 
or $\mathcal{A}$). The probability $h$ of leaving a strategy type is therefore the probability that a randomly drawn memory-1 strategy replaces a randomly drawn resident of that type. 
For the self-alternators we have

\[
h_{a}=\int_{\mathbf{q}\in\mathcal{A}}\int_{\mathbf{p}\in[0,1]^4}\rho(\mathbf{q},\mathbf{p})\mathbf{dp}\mathbf{dq}
\]
\\
where $q$ is integrated over all self-alternator strategies $\mathcal{A}$. Similarly we have

\[
h_{c}=\int_{\mathbf{q}\in\mathcal{C}}\int_{\mathbf{p}\in[0,1]^4}\rho(\mathbf{q},\mathbf{p})\mathbf{dp}\mathbf{dq}
\]
\\
for self-cooperators, where $q$ is integrated over all self-cooperator strategies $\mathcal{C}$. And

\begin{equation}
h_{d}=\int_{\mathbf{q}\in\mathcal{D}}\int_{\mathbf{p}\in[0,1]^4}\rho(\mathbf{q},\mathbf{p})\mathbf{dp}\mathbf{dq}
\end{equation}
\\
for self-defectors, where $q$ is integrated over all self-defector strategies $\mathcal{D}$. The stationary distribution of this three-state Markov chain with these transition probabilities can be readily found to give

\[
\Pi_{a}=\frac{g_a/h_a}{g_a/h_a+g_c/h_c+g_d/h_d}
\]
\\
for the probability of the system to be at an self-alternator strategy,

\[
\Pi_{c}=\frac{g_c/h_c}{g_a/h_a+g_c/h_c+g_d/h_d}
\]
\\
for the probability of the system  to be at a self-cooperator strategy, and

\[
\Pi_{d}=\frac{g_d/h_d}{g_a/h_a+g_c/h_c+g_d/h_d}
\]
\\
for the probability of the system to be at a self-defector strategy. 

As shown in Fig.~2 and Fig.~4 of the main text, the analytic expressions above for the amount of time spent at each strategy 
type, given the current payoff matrix, provide very good approximations for the actual occupancy times observed in 
Monte-Carlo simulations over all strategies, even as the payoff matrix evolves.

\subsection*{Relaxation of assumptions}
We now relax each of four assumptions made in the main text: strong selection, rapid mutations to payoffs, ``private'' mutations to payoffs, and
global mutations to strategies.
\\
\\
\textbf{Necessary conditions for memory-1 strategies to be evolutionary robustness under weak selection:}
We have so far assumed that selection is strong. However, we can relax this assumption, and consider instead the robustness of memory-1 strategies in the regime of 
weak selection, $N\to\infty$ with $N\sigma$ fixed, as in \cite{Stewart:2013fk}.  
For a population evolving under strong selection, i.e.~in the limit $N\to\infty$ with fixed $\sigma$, 
a strategy $X$ is evolutionary robust iff $s_{xx}\geq s_{xy}$ for all $Y$, i.e if no mutant is selected to invade. 
For a population evolving under weak selection, i.e.~in the limit $N\to\infty$ with $N\sigma$ fixed, even deleterious mutants may reach high frequency due to genetic drift. Therefore, in order to find the strategies that are evolutionary robust under weak selection, we must look at the probability of fixation, $\rho(\mathbf{p_x},\mathbf{q_y})$. In particular, a strategy $X$ is robust under weak selection iff $\rho(\mathbf{p_x},\mathbf{q_y})\leq1/N$ for all mutants $Y$, where $1/N$ is the probability of neutral fixation. The expression for $\rho(\mathbf{p_x},\mathbf{q_y})$ under weak selection can be Taylor expanded to give the following robustness condition: a strategy $X$ is evolutionary robust iff  

\begin{equation}
(N-2)(s_{yy}-2s_{xx}+2s_{yx}-s_{xy})>3(s_{xy}-s_{yx})
\end{equation}
\\
where $N$ is the population size \cite{Nowak:2006ly}. In this regime, we will first show that
only the three strategy types, self-alternators, self-cooperators and self-defectors, can potentially be robust, just as under strong selection. 
We then further show that under weak selection a robust strategy must 
maximize the sum of a player's score and her opponent's score -- which implies that self-defectors are never robust under weak selection.

First we derive necessary conditions for robustness. Recall that, for a resident strategy $X$ we can write

\[
s_{xx}=\kappa_{x}-\frac{\lambda_{x}}{1-\chi_{x}}(v_{cd}+v_{dc})
\]
\\
for the payoff of $X$ against itself and the payoff of a mutant $Y$ against $X$ is

\[
s_{yx}=\frac{(1-\chi_x)\kappa_x-\lambda_x(w_{cd}+w_{dc})+\chi_x((1-\chi_y)\kappa_y-\lambda_y(w_{cd}+w_{dc}))}{(1-\chi_x \chi_y)}
\]
\\
Similarly we have

\[
s_{yy}=\kappa_{y}-\frac{\lambda_{y}}{1-\chi_{y}}(v*_{cd}+v*_{dc})
\]
\\
for the payoff of $Y$ against itself and the payoff of a mutant $X$ against $Y$ is

\[
s_{xy}=\frac{(1-\chi_y)\kappa_y-\lambda_y(w_{cd}+w_{dc})+\chi_y((1-\chi_x)\kappa_x-\lambda_x(w_{cd}+w_{dc}))}{(1-\chi_x \chi_y)}
\]
\\
consider, as before, a resident strategy with $p_{cc}<1$ and $p_{dd}>0$, along with a mutation that results in a small 
change to $\kappa_y=\kappa_x+\eta$, and a small change to $\lambda_y$ so that  $(1-\chi_x)\kappa_x-\lambda_x=(1-\chi_y)\kappa_y-\lambda_y$. We then have

\[
s_{yx}=\kappa_x+\frac{\chi_{x}\eta}{1+\chi_x}(1-(w_{cd}+w_{dc}))-\frac{\lambda_x}{1-\chi_x}(w_{cd}+w_{dc})\]
\\
as well as
\[
s_{xy}=\kappa_x+\frac{\eta}{1+\chi_x}(1-(w_{cd}+w_{dc}))-\frac{\lambda_x}{1-\chi_x}(w_{cd}+w_{dc})
\]
\\
and

\[
s_{yy}=\kappa_{x}+\eta(1-(v*_{cd}+v*_{dc}))-\frac{\lambda_{x}}{1-\chi_{x}}(v*_{cd}+v*_{dc})
\]
\\
Also note that $(v*_{cd}+v*_{dc})-(v_{cd}+v_{dc})=A^*\eta$ where $A^*$ is finite and is zero if $p_{cd}=0$ and $p_{dc}=1$ or $p_{cd}=1$ and $p_{dc}=0$.
We can then write

\[
s_{xy}-s_{yx}=\eta\frac{1-\chi_x}{1+\chi_x}(1-(v_{cd}+v_{dc}))
\]
\\
and

\[
s_{xx}-s_{yx}=A\eta\frac{\lambda_x}{1-\chi_x}-\eta\frac{\chi_x}{1+\chi_x}(1-(v_{cd}+v_{dc}))
\]
\\
and

\[
s_{yy}-s_{xy}=\eta\frac{\chi_x}{1+\chi_x}(1-(v_{cd}+v_{dc}))-\eta\frac{\lambda_x}{1-\chi_x}(A^*-A)
\]
\\
where terms $\BigO{\eta^2}$ and greater have been neglected. Replacing these expressions into Eq.~16 gives

\[
\eta(N-2)\left[3\frac{\chi_x}{1+\chi_x}(1-(v_{cd}+v_{dc}))-\frac{\lambda_x}{1-\chi_x}(A^*+A)\right]>3\eta\frac{1-\chi_x}{1+\chi_x}(1-(v_{cd}+v_{dc}))
\]
\\
\\
This can always be satisfied unless $v_{cd}+v_{dc}=1$ and $A^*+A=0$, which occurs iff $p_{cd}=0$ and $p_{dc}=1$ or $p_{cd}=1$ and $p_{dc}=0$, i.e.~if the resident 
is an self-alternating strategy.
\\
\\
Similarly, we can consider mutations that change $\lambda_{x}$ by a small amount, for strategies with $0<p_{cd}<1$ and $0<p_{dc}<1$. 
The resulting payoffs following such a mutation are

\[
s_{xx}=\kappa_{x}-\frac{\lambda_{x}}{1-\chi_{x}}(v_{cd}+v_{dc})
\]
\\
for the payoff of $X$ against itself and the payoff of a mutant $Y$ against $X$ is

\[
s_{yx}=\kappa_x-\frac{\lambda_x}{1-\chi_x}(w_{cd}+w_{dc})-\eta\frac{\chi_x}{1+\chi_x^2}(w_{cd}+w_{dc})
\]
\\
Similarly we have

\[
s_{yy}=\kappa_{x}-\frac{\lambda_{x}}{1-\chi_{x}}(v*_{cd}+v*_{dc})-\eta\frac{1}{1-\chi_{x}}(v*_{cd}+v*_{dc})
\]
\\
for the payoff of $Y$ against itself and the payoff of a mutant $X$ against $Y$ is

\[
s_{xy}=\kappa_x-\frac{\lambda_x}{1-\chi_x}(w_{cd}+w_{dc})-\eta\frac{1}{1+\chi_x^2}(w_{cd}+w_{dc})
\]
\\
We can then write

\[
s_{xy}-s_{yx}=-\eta\frac{1}{1+\chi_x}(v_{cd}+v_{dc})
\]
\\
and

\[
s_{yy}-s_{xy}=\eta\frac{\lambda_x}{1-\chi_x}(A-A^*)-\eta\frac{\chi_x}{1+\chi_x^2}(v_{cd}+v_{dc})
\]
\\
where in this case $A^*=0$ if $p_{cc}=1$ or if $p_{dd}=0$. We also have

\[
s_{xx}-s_{yx}=\eta\frac{\lambda_x}{1-\chi_x}A+\eta\frac{\chi_x}{1+\chi_x^2}(v_{cd}+v_{dc})
\]
\\
\\
Replacing these expressions into Eq.~16 gives

\[
\eta(N-2)\left[\frac{\lambda_x}{1-\chi_x}(A+A^*)+3\frac{\chi_x}{1+\chi_x^2}(v_{cd}+v_{dc})\right]<3\eta\frac{1}{1+\chi_x}(v_{cd}+v_{dc})
\]
\\
which can always be satisfied unless $v_{cd}+v_{dc}=0$ and $A+A^*=0$, which occurs iff $p_{cc}=1$ or $p_{dd}=0$, i.e.~if the resident strategy is either a self-cooperator or a self-defector.
\\
\\
Thus, in total, we have shown that only self-alternators, self-cooperators and self-defectors can be robust under weak selection.
However we can also construct a strategy that will selectively replace any resident that does not achieve the maximum possible score against itself. 
To see this, consider a resident self-alternator with $p_{cd}=0$ and $p_{dc}=1$, which scores $s_{xx}=(1/2)(R(cd)+R(dc))$. If $2R(cc)>R(cd)+R(dc)$,
we can construct a mutant $Y$ with $p_{cd}=0$, $p_{dc}=1$ and $p_{cc}=1$. Such a mutant scores $s_{yx}=s_{xy}=s_{xx}$. However it also scores 
$s_{yy}=(1/2)(R(cc)+s_{xx})$ (assuming that there is an error rate $\epsilon$ in the player's execution of their strategy \cite{Fund2}). Therefore we have $s_{yy}>s_{xx}$, 
which means that $Y$ is selected to replace $X$, according to Eq.~16. A similar argument holds for any resident of the type self-alternator, self-cooperator or self-defector, 
unless $s_{xx}$ is maximum. This implies that, in fact, under weak selection only self-cooperators or self-alternators can be robust, 
since by definition self-defectors do not maximize their scores. 
\\
\\
Note that a memory-1 strategy that is robust under weak selection is robust against all opponents, regardless of their memory. Although
the robustness conditions under weak selection depend on $s_{yy}$, we have shown that in order to be robust $s_{xx}$ must be maximized. As a result, no
opponent can do better against himself than a resident robust strategy does against herself.
\\
\\
\textbf{The collapse of cooperation under weak selection:}
Sufficient conditions for a strategy to be robust under weak selection can  be found using Eqs.~5-10 along with Eq.~16. The case
$2R(cc)>R(dc)+R(cd)$, for example, in which only a subset of self-cooperators are robust, has been studied by \cite{Stewart:2013fk}. For the donation game, these conditions
reduce to

\[
\lambda>\frac{B-C}{3N}\left[N+1-(2N-1)\chi\right]
\]
\\
and

\[
\lambda>\frac{B+C}{N-2}\left[N+1-(2N-1)\chi\right]
\]
\\
Using Eq.~1 to convert back to the standard coordinate system we have

\begin{eqnarray*}
\left[3N(B+C)+(2N-1)(B-C)\right](1-p_{ab})>
\left[3N(B+C)-(2N-1)(B-C)\right]p_{ba}
\end{eqnarray*}
\\
\\
and

\[
2(N-2)(B-C)(1-p_{ab})>\left[3N(B+C)-(N-2)(B-C)\right]p_{bb}
\]
\\
Just as in the case of strong selection, the volume of robust self-cooperative strategies shrinks as the ratio of benefits to costs shrinks. 
And so this analysis predicts a collapse of
cooperation as payoffs evolve towards higher values. This behavior is indeed confirmed by Monte-Carlo simulations (Fig.~S7), 
illustrating that the collapse of cooperation occurs
under both strong and weak selection.
\\
\\
\textbf{Alternate mutation schemes:}
We have focused in the main text on a mutation scheme in which $\gamma=1.25$, so
that costs and benefits occur in the relationship $B=1.25C+k$. 
The collapse of cooperation persists, to a lesser or greater extent, when
larger or smaller values of $\gamma>1$ are considered, as shown in Fig.~S3a-c.
These payoff-mutation schemes all correpond to a tradeoff in which larger benefits
of mutual cooperation, $B-C$, are
associated with larger costs of being defected against, $C$.

We can alterantively consider values $0<\gamma<1$. In this case, as $B$ and $C$
increase, the benefit for mutual cooperation, $B-C$, decreases.  And so larger
benefits of mutual cooperation are no longer associated with larger costs of being
defected against. When strategies and payoffs co-evolve under this mutation scheme
selection leads to \textit{decreasing} values of $B$ and $C$, until $C$ reaches
zero (so that there is no longer a Prisoner's Dilemma). As might be expected,
there is a collapse of defection in this case, with self-defectors replaced by
self-cooperators as payoffs evolve (Fig.~S4). 

Finally, in the limiting
case in which $B$ can increase indefinitely and $C$ remains fixed, self-cooperators become more
successful as $B$ evolves and there is no collapse of cooperation (Fig.~S3d).
\\
\\
\textbf{Public mutations to payoffs:}
We have assumed that mutations to payoffs affect only the individual carrying the
mutation -- so called ``private" mutations. This assumption can be relaxed in a number of ways, to
reflect the fact that a social interaction is occurring. One natural alternative
is to assume that each player sets their opponent's costs or benefits. In the first
case, when players set their opponent's cost $C$, the payoffs received by  player
$X$ facing opponent $Y$ are $R_x(cc)=B_x-C_y$, $R_x(cd)=-C_y$, $R_x(dc)=B_x$
and $R_x(dd)=0$.
Under this payoff mutation scheme we again find that higher payoffs evolve, precipitating
the collapse of
cooperation (Fig.~S5). However, in the second case, when players set their
opponent's benefits $B$, the payoffs received by  player $X$ facing 
opponent $Y$ are $R_x(cc)=B_y-C_x$, $R_x(cd)=-C_x$, $R_x(dc)=B_y$ and $R_x(dd)=0$, 
and we find evolution towards increased self-cooperation, despite decreasing rewards for
mutual cooperation, $B-C$ (Fig.~S6).
\\
\\
\textbf{Slow mutations to payoffs:}
In the main text we assumed that mutations to payoffs and mutations to strategies occur at equal rates. This assumption can be relaxed to allow for the 
scenario in which mutations to payoffs are relatively more rare. As shown in Fig.~S8 the collapse of cooperation persists even when mutations 
to payoffs are relatively rare.
\\
\\
\textbf{Local mutations to strategies:}
In the main text we assumed that mutations to strategies are global, such that a mutant was drawn uniformally from the space of all memory-1 strategies. 
This assumption can be altered to consider the 
scenario in which mutations to strategies increase or decrease each element of a memory-1 strategy, $\mathbf{p}$, by a small amount $\Delta$, 
with the constraint that mutant probabilities lie in the range $[0,1]$. As shown in Fig.~S9 the collapse of cooperation persists when mutations
to strategies are local.

\bibliographystyle{nature.bst}

\clearpage

\section*{Supplementary figures}

\begin{figure*}[h] \centering
\subfigure{\includegraphics[scale=0.3]{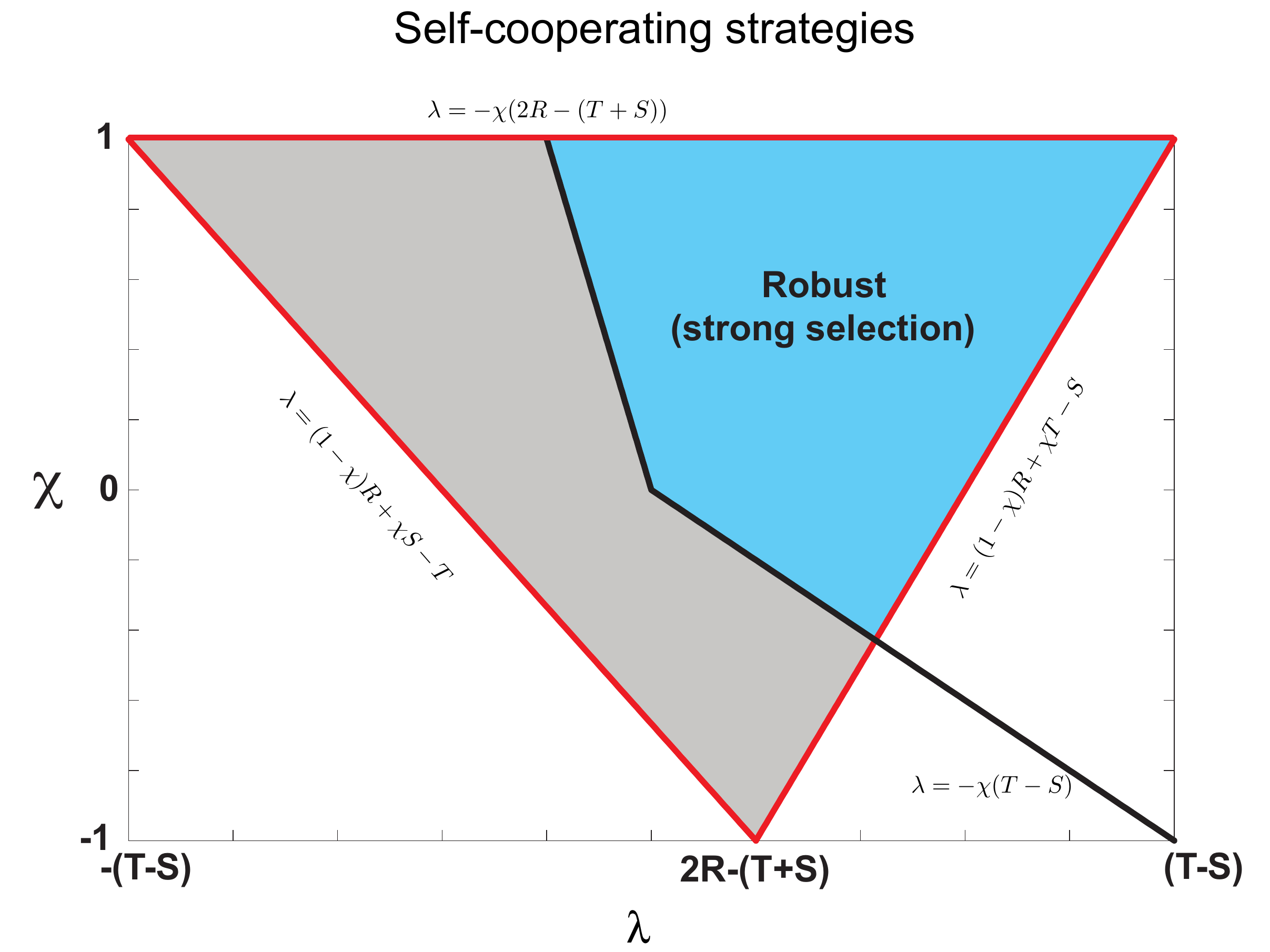}}
\subfigure{\includegraphics[scale=0.31]{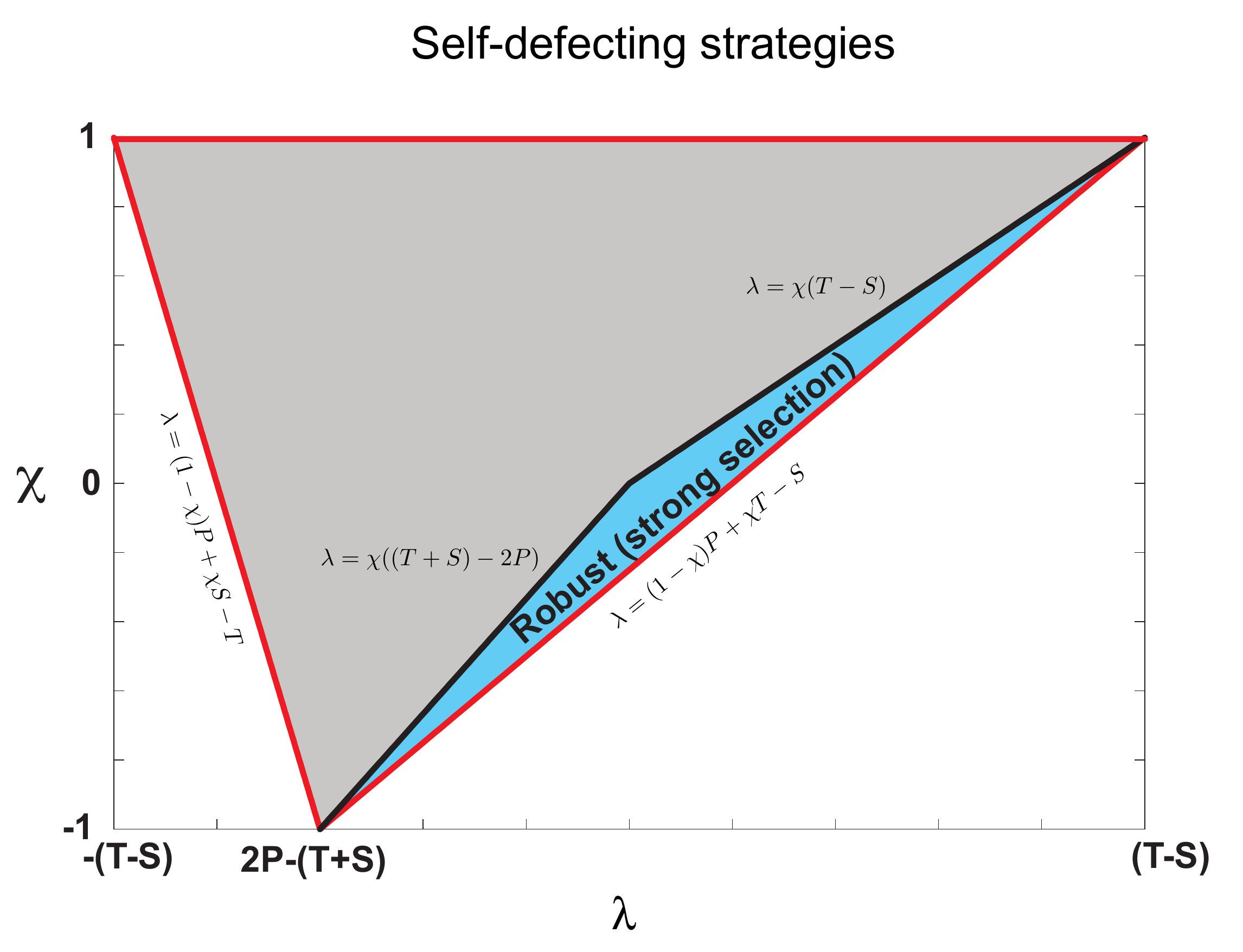}}
\subfigure{\includegraphics[scale=0.31]{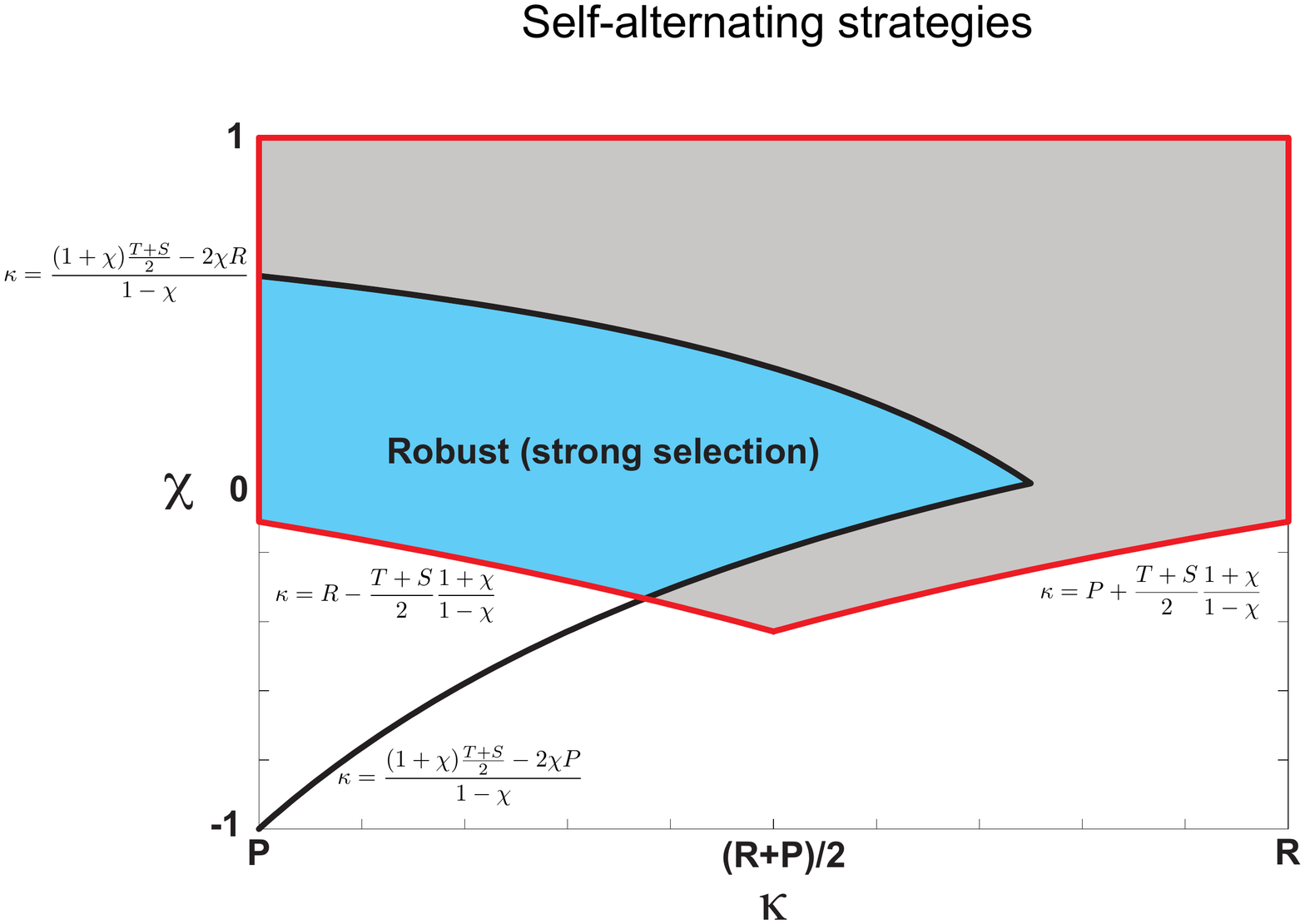}} 
\caption*{\small Figure S1 -- Confirmation by Monte-Carlo simulation of analytical conditions for evolutionary robustness of strategies.  
For each of the three strategy types, self-cooperators ($p_{cc}=1$), self-defectors ($p_{dd}=0$), and self-alternators ($p_{cd}=0$ and $p_{dc}=1$), we compare
analytic expression for evolutionary robustness (black lines) with numerical calculations of robustness (light blue regions).
Coordinates
$(\kappa,\chi)$ for the self-alternating strategies and $(\lambda,\chi)$ for self-cooperators and self-defectors were sampled in regular intervals of 0.01 within the
space of all feasible strategies (outlined in red).
For each sampled pair of co-ordinates $(\lambda,\chi)$ we also sampled $10^3$ associated values of $\phi$, ranging from $\phi\to0$ to the 
maximum feasible $\phi$.
To determine numerically whether a focal strategy $X=(\lambda,\chi,\phi)$ is robust we computed the longterm
payoffs $s_{xx}$, $s_{yy}$, $s_{xy}$ and $s_{yx}$ against $10^6$ opponent strategies, $Y$, drawn uniformly from all memory-1 
strategies. A focal strategy $X$ was designated as robust if no strategy $Y$ was found with a score $s_{yx}>s_{xx}$. 
Parameters are $N=100$, $\sigma=10$, $R(cc)=R=3$, $R(cd)=S=0$,
$R(dc)=T=5$ and $R(dd)=P=1$.  }
\end{figure*}

\clearpage

\begin{figure*}[h] \centering 
\includegraphics[scale=0.66]{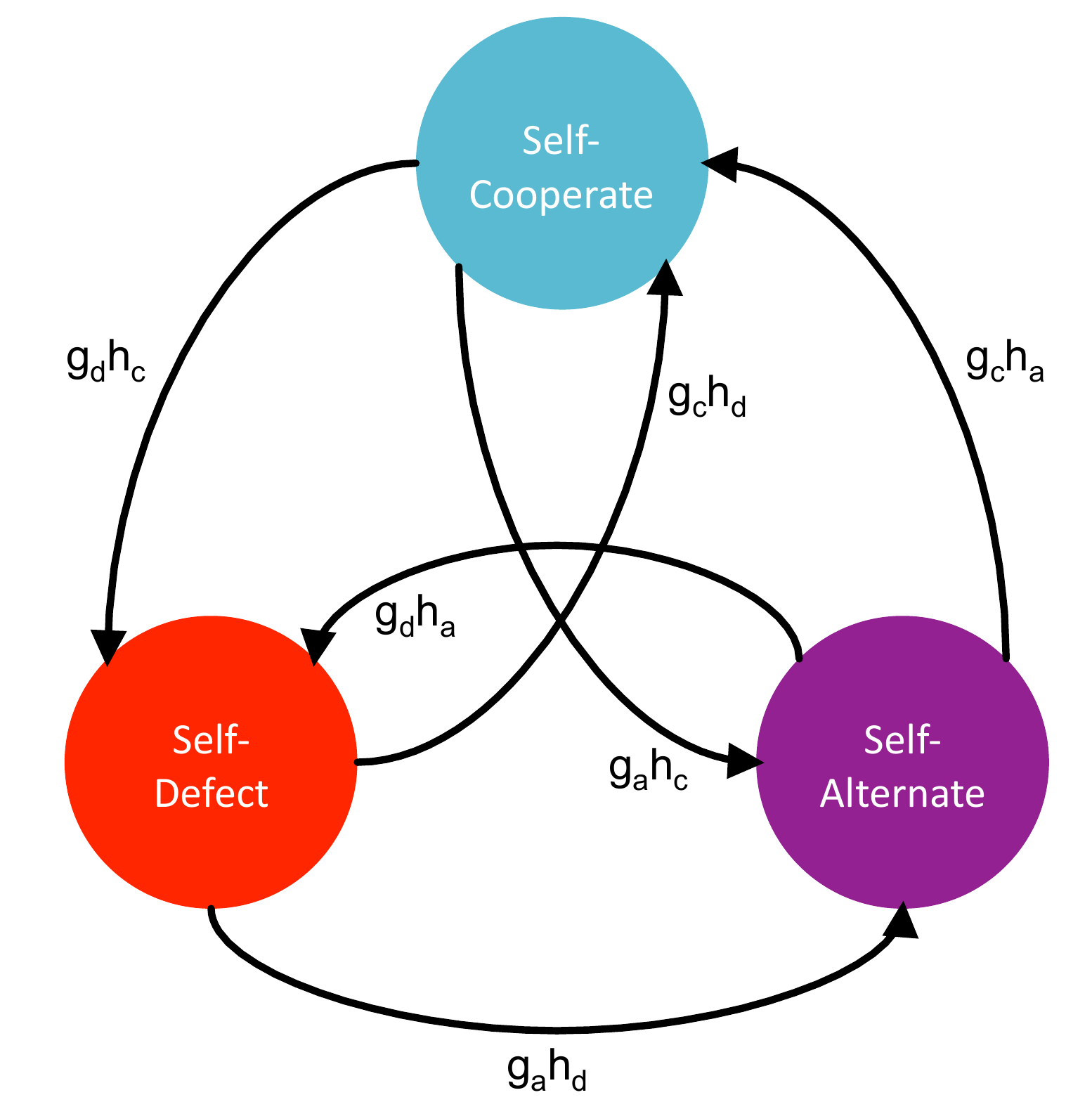}
\caption*{\small Figure S2 -- A simplified, three-state Markov chain 
to describe evolution of strategies in two-player games. The transition rates are as given by 
Eqs.~14-15. In this simplified model we assume that the time spent away from 
these three strategy 
types can be neglected. 
This approximation is supported by simulations on the full space of strategies, 
which indicate that such populations occupy one of these three strategy types
$>97\%$ of the time.
}
\end{figure*}

\clearpage



\begin{figure*}[h] \centering 
\includegraphics[scale=1.0]{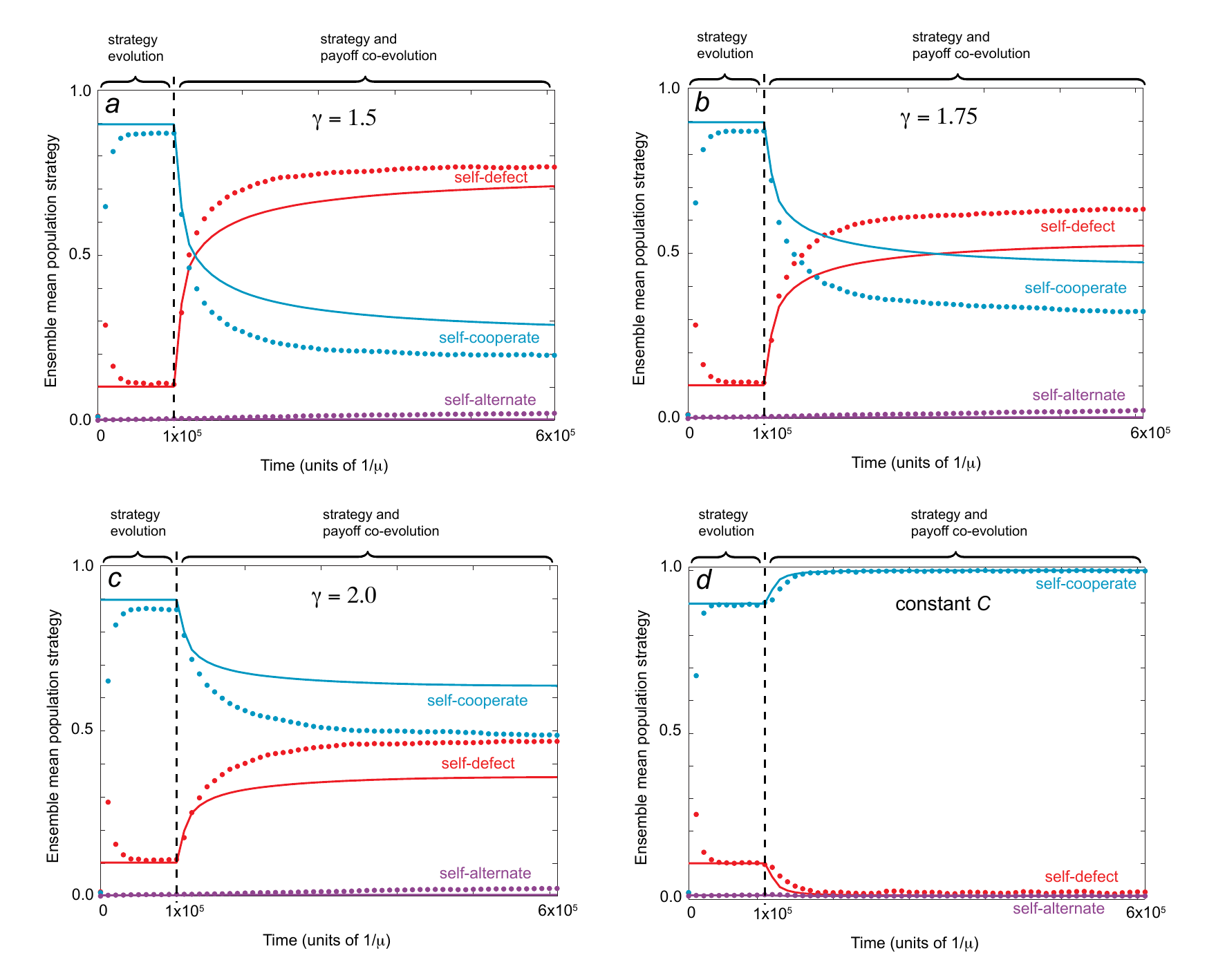}
\caption*{\small Figure S3 -- The collapse of cooperation in the Prisoner's
Dilemma under different mutation schemes.  We simulated populations under weak mutation,
proposing both mutant strategies and mutant payoffs at equal rates, $\mu/2$.
Mutations to strategies were drawn 
uniformly from the full space of memory-1 strategies.
Mutations to payoffs were drawn so that
increasing benefits of cooperation incur increasing costs: mutations perturbing
the benefit $B$ by $\Delta$ were drawn
uniformly from the range $\Delta\in\left[-0.1,0.1\right]$, with the corresponding
change to cost $C$ chosen to enforce the relationship $B=\gamma C + k$ with (a) $\gamma=1.5$, (b) $\gamma=1.75$, (c) $\gamma=2.0$ or (d) 
allowing $B$ to evolve with fixed $C=1$.
Evolution was modelled according to an imitation process under weak mutation \cite{Sig2,Stewart:2013fk,Traulsen:2006zr}.
Self-cooperative strategies are
initially robust and dominate the population, but they are quickly replaced by
self-defectors as payoffs evolve.
Dots indicate the proportion of $10^5$ replicate populations, at each time point, within distance
$\delta=0.01$ of the three strategy types self-cooperate, self-defect, and self-alternate. 
Lines indicate analytic predictions for the frequencies of these strategy
types, which depend upon the corresponding volumes of robust strategies.
Simulations were run until each population experienced $5\times 10^5$ mutations.  Populations of size $N=100$ were initiated with $B=3$ and $C=1$, and
evolved under selection strength $\sigma=1$ (strong
selection). }
\end{figure*}

\clearpage

\begin{figure*}[h] \centering 
\includegraphics[scale=1.0]{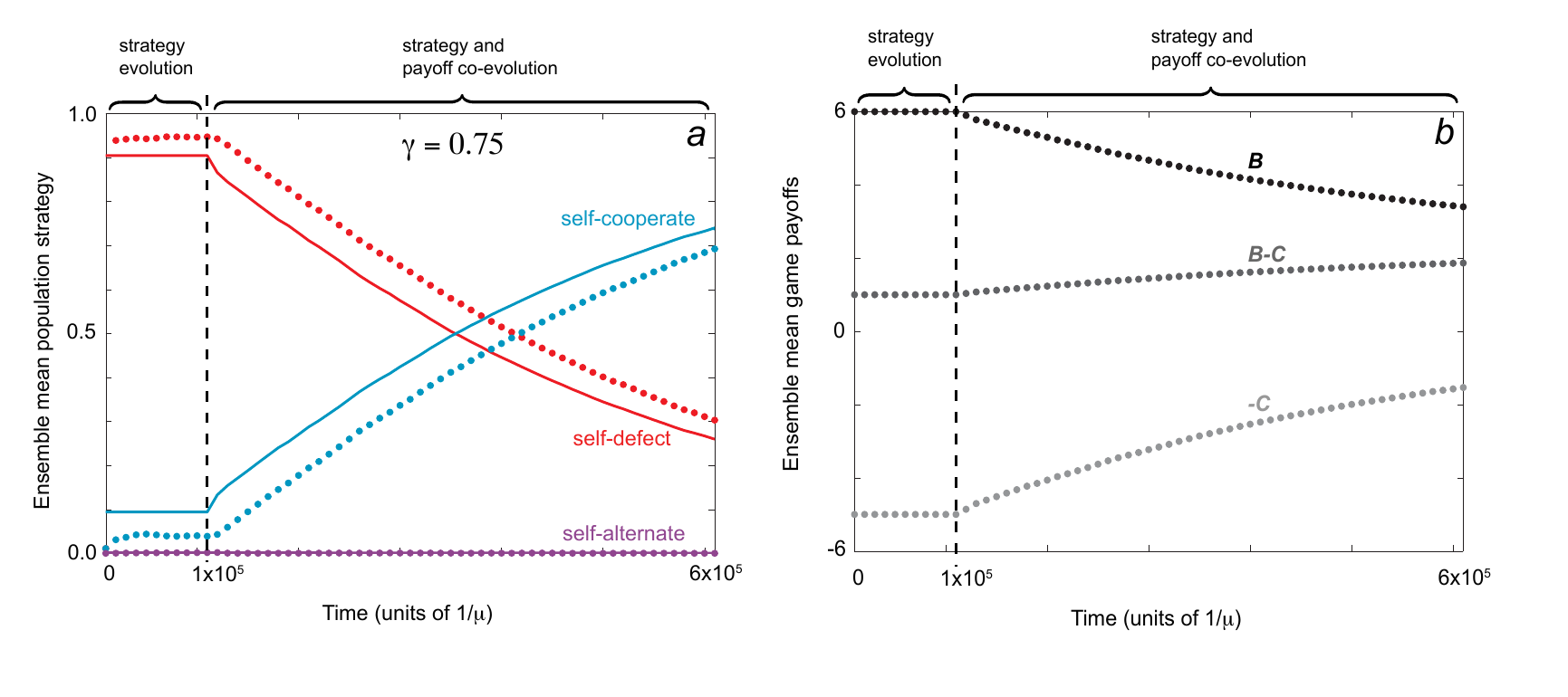}
\caption*{\small Figure S4 -- Co-evolution of strategies and payoffs when 
$B$ and $C$ are allowed
to evolve with $\gamma=0.75$, so that the payoff for mutual cooperation, $B-C$, increases as $B$ and $C$ decrease. 
(a) Populations were initialized at $B=6$ and $C=5$, under which self-defect dominates. Once strategies and payoffs start to co-evolve, 
self-cooperate begins to increase and eventually comes to dominate. (b) Benefits
$B$ and costs $C$ evolve towards lower values.  
We simulated populations under weak mutation as in Fig.~2a. Lines indicate analytic predictions for the frequencies of these strategy
types, which depend upon the corresponding volumes of robust strategies.
Simulations were run until each population had experienced $5\times 10^5$ mutations.  Populations of size $N=100$ were initiated with $B=3$ and $C=1$, and
evolved under selection strength $\sigma=1$ (strong
selection).
}
\end{figure*}

\clearpage

\begin{figure*}[h] \centering 
\includegraphics[scale=0.4]{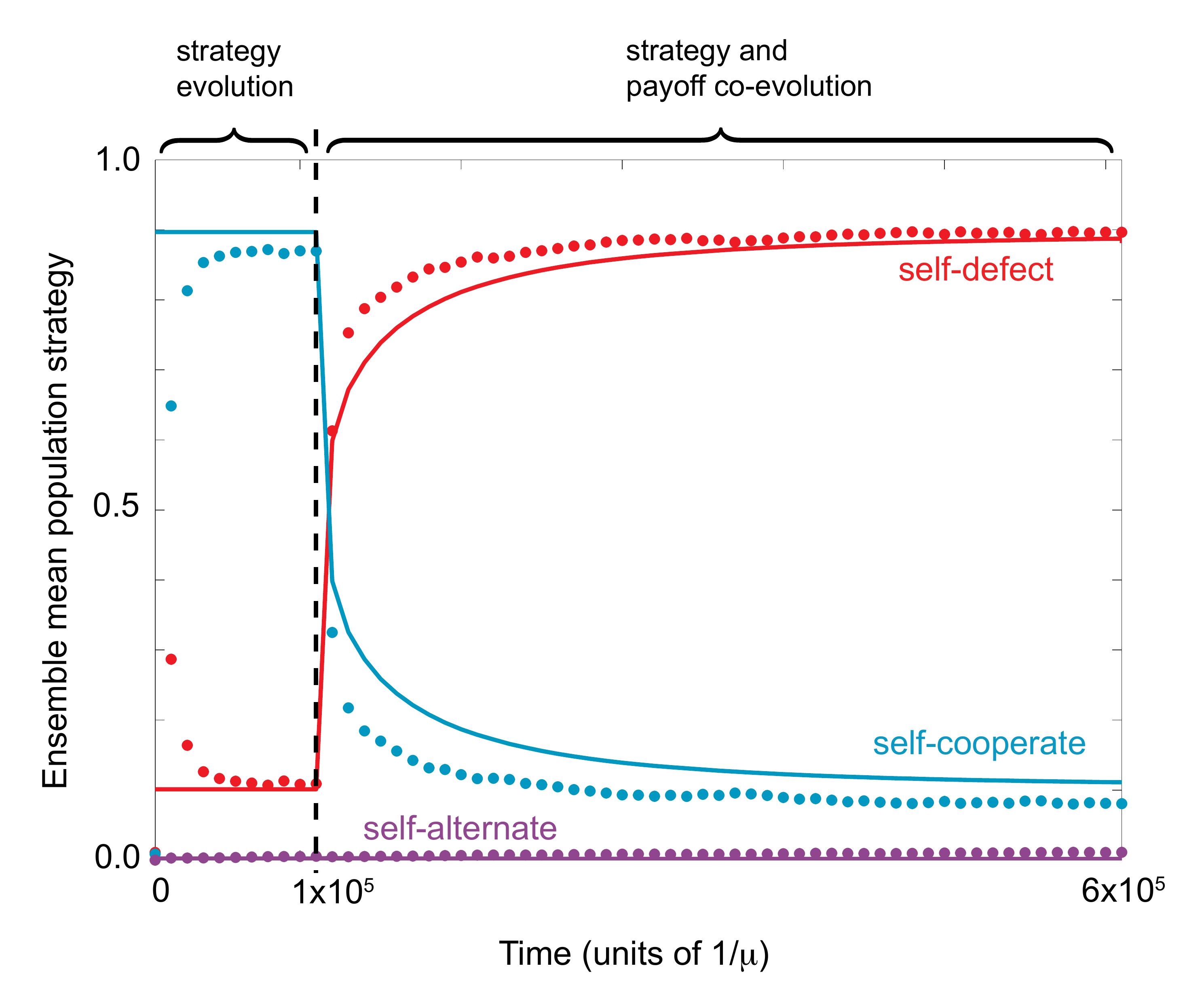}
\caption*{\small Figure S5 --  Public mutation and the collapse of cooperation in the Prisoner's Dilemma. 
We simulated populations under weak mutation, as in Fig. 2a, except that 
mutations to $C$ are ``public'' in the sense that the cost of an interaction borne by a player depends on the genotype of her opponent, 
as described
in the supplementary text.}
\end{figure*}

\clearpage

\begin{figure*}[h] \centering 
\includegraphics[scale=1.0]{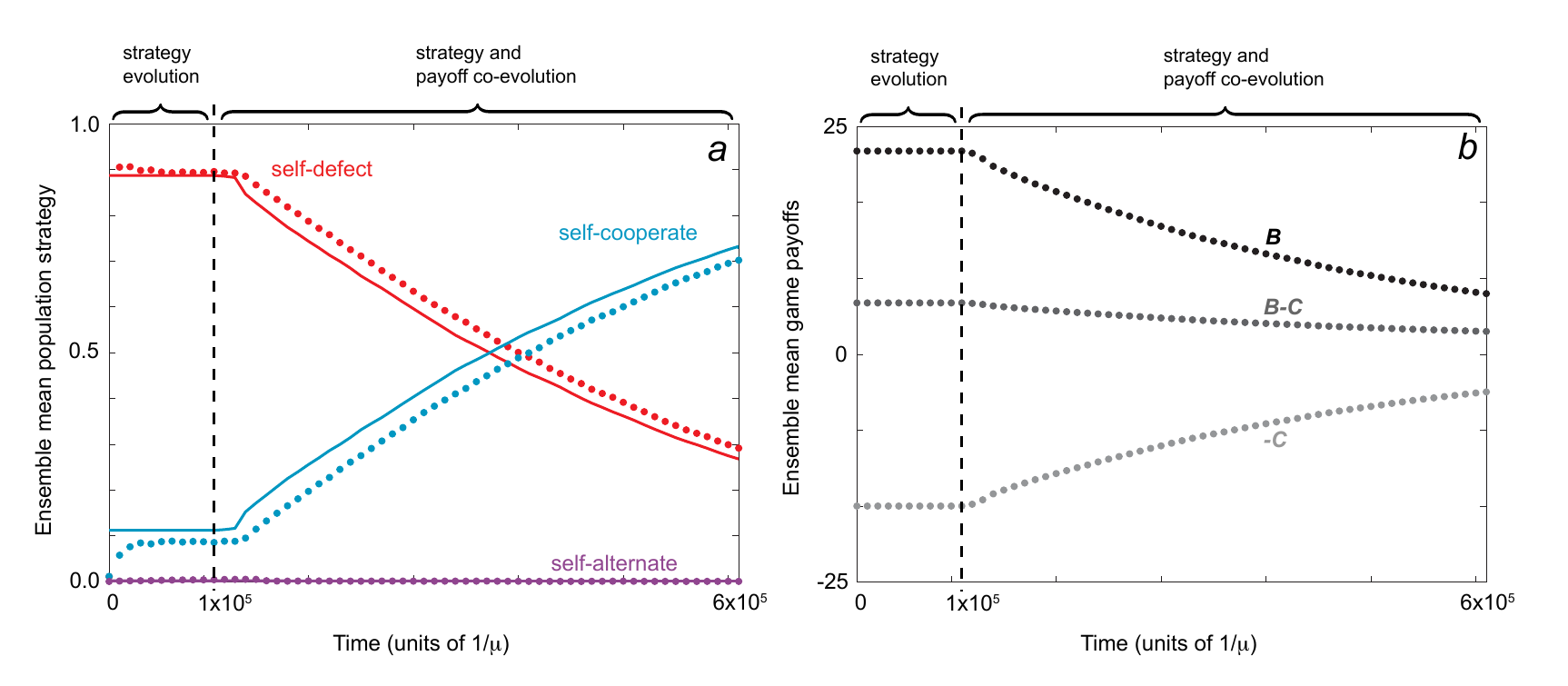}
\caption*{\small Figure S6 --  Public mutation and the collapse of cooperation in the Prisoner's Dilemma. 
(a) We simulated populations under weak mutation, as in Fig. 2a, except that 
mutations to $B$ are ``public'' in the sense that the benefit of recieved by a player depends on the genotype of her opponent, as described
in the supplementary text. In this case populations evolve away from self-defect
and towards self-cooperate. However (b) this evolution comes at the expense of decreasing
benefits for cooperation.}
\end{figure*}

\clearpage

\begin{figure*}[h] \centering 
\includegraphics[scale=0.4]{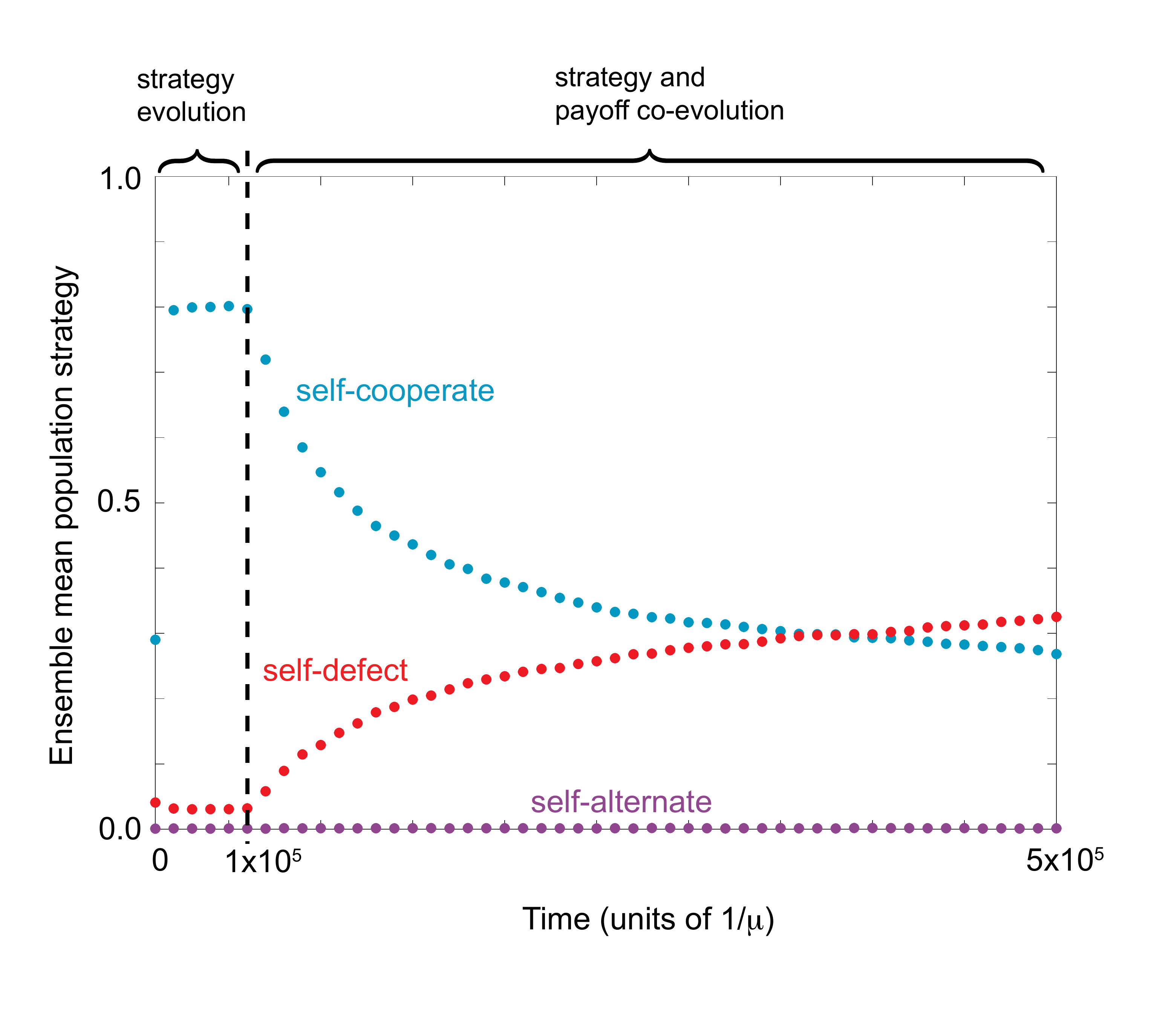}
\caption*{\small Figure S7 --  The collapse of cooperation in the Prisoner's Dilemma under weak selection.  
We simulated populations under weak mutation as in Fig. 2a, except with $N=100$ and $\sigma=0.01$ (weak selection).
Self-cooperative strategies are
initially robust and dominate the population, but they are quickly replaced by
self-defectors as payoffs evolve. 
}
\end{figure*}

\clearpage

\begin{figure*}[h] \centering 
\includegraphics[scale=0.4]{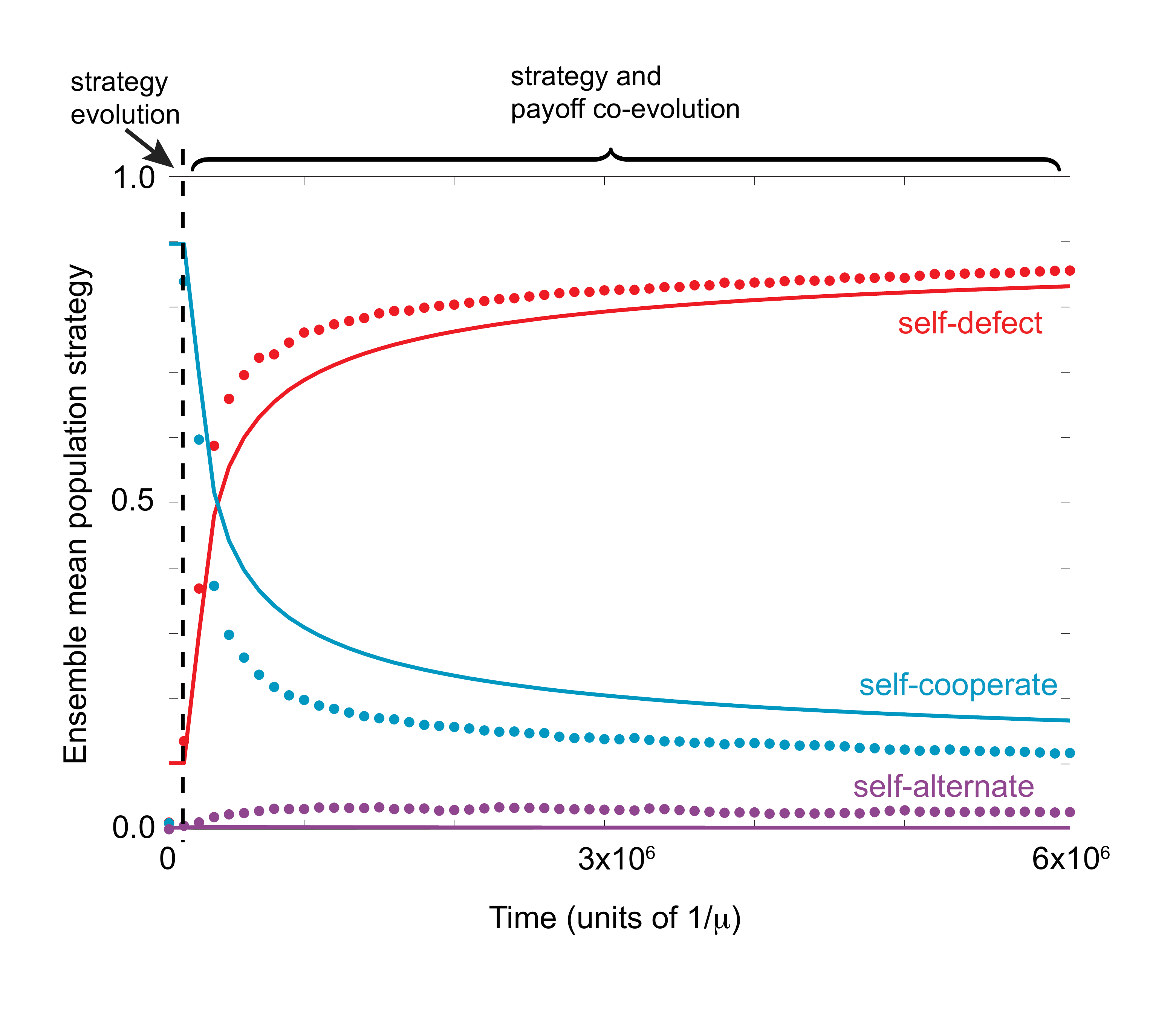}
\caption*{\small Figure S8 -- Slow mutations to payoffs and the collapse of cooperation in the Prisoner's Dilemma.  
We simulated populations under weak mutation as in Fig. 2a, except that mutations altering strategies occur at
$10^3$-times the rate of mutations altering payoffs.
Self-cooperative strategies are
initially robust and dominate the population, but they are quickly replaced by
self-defectors as payoffs evolve. 
}
\end{figure*}

\clearpage

\clearpage

\begin{figure*}[h] \centering 
\includegraphics[scale=0.4]{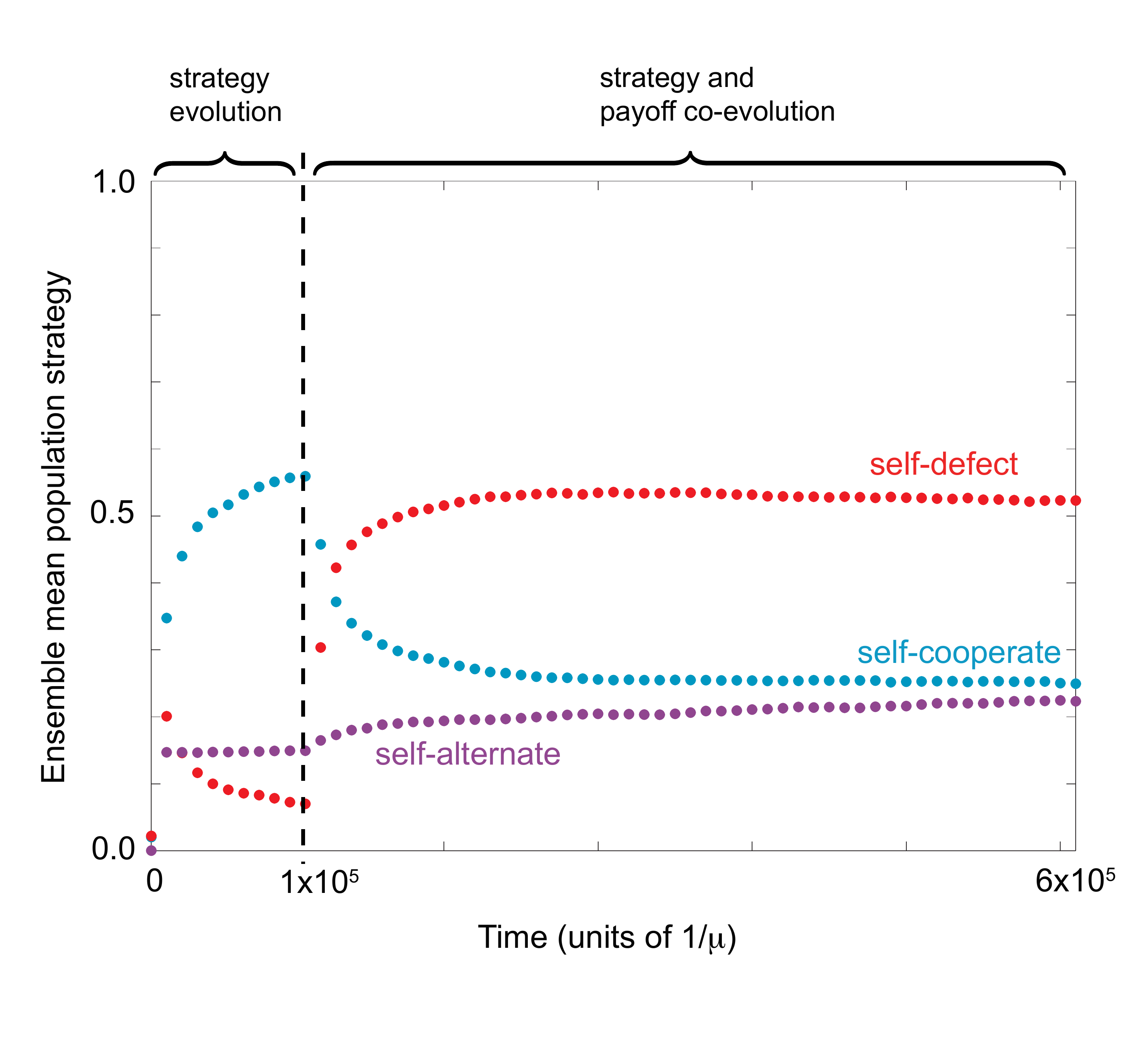}
\caption*{\small Figure S9 -- Local mutations to payoffs and the collapse of cooperation in the Prisoner's Dilemma.  
We simulated populations under weak mutation as in Fig. 2a, except that mutations altering strategies are now ``local'' so that mutations perturbing
each of the four probabilities $(p_{cc},p_{cd},p_{dc},p_{dd}$ by an amount $\Delta$ were drawn
uniformly from the range $\Delta\in\left[-0.01,0.01\right]$.
Self-cooperating strategies are
initially robust and dominate the population, but they are quickly replaced by self-defectors as payoffs evolve. }
\end{figure*}

\clearpage

\bibliographystyle{nature.bst}

\bibliographystyle{nature.bst}

\end{document}